\documentclass[12pt]{article}
\usepackage{a4,epsf,amsmath,amssymb,axodraw}
  \newcommand{\ccaption}[2]{
    \begin{center}
    \parbox{0.85\textwidth}{
      \caption[#1]{\small{#2}}
      }
    \end{center}
    }
\newcommand{\beq}{\begin{equation}}
\newcommand{\eeq}{\end{equation}}
\newcommand{\beqa}{\begin{eqnarray}}
\newcommand{\eeqa}{\end{eqnarray}}
\newcommand{\nn}{\nonumber}
\newcommand{\Eqn}[1]{Eq.~(\ref{#1})}
\newcommand{\Eqns}[2]{Eqs.~(\ref{#1}) and (\ref{#2})}
\newcommand{\ol}{\overline}
\newcommand{\pab}{\bar{\partial}}
\newcommand{\thb}{\bar{\theta}}
\newcommand{\zeb}{\bar{\zeta}}
\newcommand{\xib}{\bar{\xi}}
\newcommand{\psib}{\bar{\psi}}
\newcommand{\chib}{\bar{\chi}}
\newcommand{\phib}{\bar{\phi}}
\newcommand{\lambdab}{\bar{\lambda}}
\newcommand{\sigmab}{\bar{\sigma}}
\newcommand{\ad}{{\dot\alpha}}
\newcommand{\bd}{{\dot\beta}}
\newcommand{\mat}[4]{\left(\begin{array}{cc} #1 & #2 \\ #3 & #4
                           \end{array}\right)} 

\begin{document}
\thispagestyle{empty}
\begin{flushright}
\phantom{March 2009}
\end{flushright}
\vspace*{3cm}
\begin{center}
{\bf \Large ABC of SUSY}
\\
\vspace{2cm}
{\sc Adrian Signer}\\[2em]
{\sl Institute for Particle Physics Phenomenology,\\ 
Durham University,
Durham DH1~3LE, UK}
\end{center}
\vspace{2ex}
\begin{abstract}
This article is a very basic introduction to supersymmetry. It
introduces the two kinds of superfields needed for supersymmetric
extensions of the Standard Model, the chiral superfield and the vector
superfield, and discusses in detail how to construct supersymmetric,
gauge invariant Lagrangians. The main ideas on how to break
supersymmetry spontaneously are also covered. The article is meant to
provide a platform for further reading.
\end{abstract}

\newpage
\setcounter{page}{1}
\setcounter{footnote}{0}
\parskip =0.2cm

\hyphenation{
coun-ter-term
}

\catcode`\@=11
\@addtoreset{equation}{section}


\section{Introduction \label{sec:intro}}

This is neither a review article, nor a summary of supersymmetry.
There are already many excellent reviews available. The standard
reference for a comprehensive introduction and review of supersymmetry
has been written by Martin~\cite{Martin:1997ns}. Recently, an
introduction with applications to particle theory has also been
written by Peskin~\cite{Peskin:2008nw} and there are earlier articles
of Olive~\cite{Olive:1999ks} and Drees~\cite{Drees:1996ca}, the latter
with an extended discussion of quadratic singularities . The Physics
Reports of Haber and Kane~\cite{Haber:1984rc} and
Nilles~\cite{Nilles:1983ge} are early review articles about
supersymmetry. The former contains a comprehensive discussion of the
minimal supersymmetric extension of the Standard Model (MSSM), the
latter includes supergravity. An introduction including material for
$N>1$ supersymmetry can be found in the Tasi lecture notes of
Lykken~\cite{Lykken:1996xt}. An up-to-date view on breaking
supersymmetry is given in the lecture notes of Dine~\cite{Dine:2009gy}
or Intriligator and Seiberg~\cite{Intriligator:2007cp}. Needless to
say that this list is by no means exhaustive or in any way selective.

As the title suggests, this article is meant to guide the reader
through the first few steps of understanding susy. Thus it is for
those who have a first go at susy or usually get stuck somewhere
between page~2 and page~5 of other introductions and reviews. The hope
is that after reading this article the other articles are easier to
understand. Accordingly, this article stops where all the others begin
in earnest. In particular it does not contain any serious applications
to collider physics or cosmology nor does it cover any developments of
the past few years or anything beyond $N=1$ susy. It only covers the
very basic concepts of global $N=1$ susy, but hopefully does so in
more detail than the above mentioned articles.

The article assumes a basic understanding of field theory and gauge
theory and is meant to provide an as direct as possible path to
writing down the MSSM. At the same time it aims to be precise in that
nothing essential is left out or swept under the rug. In the main text
the basic ideas are given and illustrated. We start in
Section~\ref{sec:sym} with a discussion of symmetries and the
extension of the Poincar\'e symmetry to include susy. In
Section~\ref{sec:tech} the minimal amount of technicalities needed are
covered, Weyl spinors (which we use throughout) and Grassmann
variables. Section~\ref{sec:superspace} introduces the concepts of
superspace and superfields. These will turn out to be indispensable in
Section~\ref{sec:Lagrangians} which is the main section and discusses
the construction of susy theories. This section concludes with writing
down the unbroken MSSM after which we turn to breaking susy in
Section~\ref{sec:Breaking}. The basic possibilities to break susy
spontaneously and their problems in realistic applications are
discussed and the notion of soft breaking is explained. This is where
we stop with our ABC of SUSY and leave the reader to make the steps
from D to Z with the help of other articles. It should be possible to
follow through the main text without delving into the gory details of
conventions and indices. However, for a full understanding these 
details are required. For the reader willing to get his/her hands
dirty, the conventions used in this article are given in
Appendix~\ref{App:Conventions}. Finally, Appendix~\ref{app:sample}
presents some sample calculations whose results are used in the main
text. These details are often not available in other articles and
hopefully provide some help in understanding the technicalities.

\section{(Super)Symmetries \label{sec:sym}}

A symmetry is a group of transformations that leaves the Lagrangian
invariant. Two of the reasons why symmetries are very important are:
first according to the Noether theorem, with each continuous symmetry
we can associate a conserved quantity and second and even more
importantly, nature seems to respect many of them. A continuous
symmetry is one that depends continuously on one or several
parameters. As an example consider rotations and space
translations. To determine a three dimensional rotation completely we
need three parameters (angles) which we will denote by
$\vec\vartheta$. The parameters of the translation are denoted by
$\vec{a}$. Under such a transformation
\begin{equation}
\vec{x} \to \vec{x}\,' = R(\vec\vartheta) \cdot \vec{x}+\vec{a}
\label{rottrsf}
\end{equation}
where $R$ is a $3\times 3$ rotation matrix depending on
$\vec{\vartheta}$ and $R(\vec 0) = 1$. In a quantum mechanical system,
under such a transformation a state $\psi(\vec{x})$  transforms as
\begin{equation}
\psi(\vec{x}) \to \psi'(\vec{x}) = 
  e^{-i\,\vec{a}\cdot\vec{P}}\, 
  e^{- i\, \vec{\vartheta}\cdot \vec{J}} \psi(\vec{x})
\label{rotstate}
\end{equation}
where $J_i$ and $P_i$, $i\in\{1,2,3\}$ are called the generators of
the rotations and translations respectively. The explicit form of the
generators depends on the precise nature (spin) of the state but in
any case they satisfy the familiar commutation relations
\begin{eqnarray}
\big[P_i, P_j\big] &=& 0 \label{PPalg} \\
\big[J_i, J_j\big] &=& i\, \epsilon_{ijk}\, J_k
\label{rotalg} \\
\big[P_i, J_j\big] &=& i\, \epsilon_{ijk}\, P_k
\label{PJalg}
\end{eqnarray}
The remarkable fact is that nature respects rotational and
translational symmetry, i.e. the Lagrangian of any fundamental theory
has to be invariant under \Eqn{rottrsf}. This is a crucial help in
constructing theories that have a chance of being realized in nature.

This is all fine and good, but in fact we know we can do better.  We
can enlarge the symmetry group. The symmetry group that lies at the
heart of every Quantum Field Theory (QFT) is the Poincar\'e group
consisting of Lorentz transformations (LT) and translations
\begin{equation}
x^\mu \to x'^\mu = x^\mu + \omega^{\mu\nu} x_\nu + a^\mu
\label{Poincare}
\end{equation}
where $x^\mu = (t, \vec{x})$ denotes the coordinates in Minkowski
space-time.  To specify completely an arbitrary Poincare
transformation, we need six Lorentz parameters (three boost parameters
$\vec{\varphi}$ and three rotation angles $\vec{\vartheta}$), written
in terms of an antisymmetric tensor of rank two,
$\omega^{\mu\nu}=-\omega^{\nu\mu}$, as well as four translation
parameters $a^\mu$. Thus, the LT involves six generators, three for
rotations and three for boosts. They are written in terms of an
antisymmetric tensor $M^{\rho\sigma} = -M^{\sigma\rho}$, where the
Lorentz labels $\rho, \sigma$ play the role of the label $i$ in $J$
above. The translations require four generators $P^\rho$, one for each
direction. The quantities $P^\rho$ and $M^{\rho\sigma}$ correspond to
the 4-momentum and the generalized angular momentum.

The explicit form of the generators depends on the nature of the field
they act on. For a spin 1/2 field e.g. we have
\begin{equation}
P^\rho = i\, \partial^\rho\,;\qquad
M^{\rho\sigma} = i(x^\rho  \partial^\sigma -  x^\sigma\partial^\rho) + 
\frac{i}{4} [\gamma^\rho,\gamma^\sigma]\, ;
\label{MPrep}
\end{equation}
whereas for a scalar field, the last term in $M^{\rho\sigma}$,
corresponding to the spin, is absent. The transformation of an
arbitrary classical field $\Phi$ under \Eqn{Poincare} can now be
written as
\begin{eqnarray}
{\rm translations}&:&  \Phi(x) \to \Phi'(x) = e^{i\, a^\rho P_\rho} \Phi(x) 
\label{translation} \\
{\rm LT}&:& \Phi(x) \to \Phi'(x) = e^{\frac{i}{2}\,
  \omega^{\rho\sigma} M_{\rho\sigma}} \Phi(x)
\label{LT}
\end{eqnarray}
The factor $1/2$ in \Eqn{LT} is conventional and compensates for the
fact that in summing over $\rho$ and $\sigma$ we count every term
twice due to the antisymmetry. The dependence on the nature of the
field $\Phi$ is only implicit in the representation to be used for the
generators. Note that \Eqns{translation}{LT} contain \Eqn{rotstate} as
a special case.

Finally, we can look at the algebra of the Poincar\'e group, i.e. the
commutation relations between the various $P^\rho$ and
$M^{\rho\sigma}$.  They can be obtained by using \Eqn{MPrep} and
$[x^\rho,P^\sigma] = -i\, g^{\rho \sigma}$ and read
\begin{eqnarray}
\left[P^\rho,P^\sigma\right] &=& 0 \label{algPP}\\
\left[P^\rho,M^{\nu\sigma}\right] &=& 
i (g^{\rho\nu} P^\sigma - g^{\rho\sigma} P^\nu) \label{algPM} \\
\left[M^{\mu\nu}, M^{\rho\sigma}\right]&=& -i (
g^{\mu\rho} M^{\nu\sigma}+g^{\nu\sigma} M^{\mu\rho}-
g^{\mu\sigma} M^{\nu\rho}-g^{\nu\rho} M^{\mu\sigma})
\label{algMM}
\end{eqnarray}
Note that as for Eqs.~(\ref{PPalg})--(\ref{PJalg}),
Eqs.~(\ref{algPP})--(\ref{algMM}) are independent on the nature/spin
of the fields, i.e. on whether or not we include the second term of
$M^{\rho\sigma}$ in \Eqn{MPrep}.  What is important for us is that all
generators mix, in particular, according to \Eqn{algPM}, the
translations and LT are linked together.

Let us pause for a moment to consider what we have done in going from
the symmetry under \Eqn{rottrsf} to \Eqn{Poincare}. We have increased
the symmetry group from 6 generators to 10 generators. In doing so, we
have also increased the number of coordinates that are involved in the
transformations from 3 in $\vec{x}$ to 4 in $x^\mu$. Note also, that
the ``new'' generators such as $M^{0i}$ etc. mix in a non-trivial way
with the ``old'' ones such as $J_i$. The latter are latent in
$M^{ij}$.

Since nature respects Poincar\'e symmetry, it is natural to ask,
whether the symmetry can be extended even further. The answer is
obviously yes, since this is precisely what is done in gauge
theories. For a certain gauge group, say $SU(N)$ we add generators
$T^a$ with $a\in\{1\ldots N^2-1\}$. A finite gauge transformation is
then specified by $N^2-1$ parameters $\omega^a$ and is written as
$\exp(i \omega^a\, T^a)$. However, such an extension is called {\it
  trivial} because the ``new'' generators all commute with all of the
``old'' generators 
\begin{eqnarray}
\label{algTT}
\left[T^a, T^b\right] &=& i\, f^{abc} T^c \\
\left[T^a, P^\rho\right] &=& 0 \label{algTP} \\
\left[T^a, M^{\rho\sigma}\right] &=& 0 \label{algTM} 
\end{eqnarray}
where $f^{abc}$ are the structure constants of the gauge group.  This
means that the extended symmetry group is a direct product of
the Poincar\'e group with a gauge (or internal symmetry) group.

Such extensions of the Poincar\'e group are very successful in
describing particle interactions, but not really what we are
after. The question is whether we can extend the Poincar\'e group in a
non-trivial way, such that the new generators mix with $P^\rho$ and/or
$M^{\rho\sigma}$. The answer to this question is given by the {\it
  Coleman-Mandula no-go theorem}~\cite{Coleman:1967ad}, which states
that any symmetry compatible with an interacting relativistic QFT is
of the form of a direct product of the Poincar\'e algebra with an
internal symmetry, such as gauge symmetry.

This would be the end of this article if it was not for the fact that
for every no-go theorem there is usually a way around. In the proof of
the Coleman-Mandula theorem there was an implicit assumption that only
bosonic generators are involved. A bosonic generator is a generator
that transforms a bosonic (fermionic) state into another bosonic
(fermionic) state.  All generators $P^\rho, M^{\rho\sigma}$ and $T^a$
are obviously bosonic since they do not change the spin of the state
they act on. What if we allow {\it fermionic generators}, more
precisely generators that change the spin of the state by 1/2?  It is
clear that such a generator has to have a spinor label $\alpha$ for
if it acts e.g. on a scalar (spin 0) state it generates an spin 1/2
state. Thus, denoting the fermionic generator by $Q_\alpha$ we have
\begin{equation}
Q_\alpha |{\rm bos}\rangle =  |{\rm ferm}\rangle_\alpha \, ;\qquad
Q_\alpha |{\rm ferm}\rangle^\alpha =  |{\rm bos}\rangle \, ;
\label{FermGen}
\end{equation}
We will be working with Weyl spinors throughout. To represent a Dirac
spinor with four components, we need two Weyl spinors (see
Section~\ref{sec:tech}) which are conventionally denoted by $Q_\alpha$
and $\bar{Q}_\bd$ with $\alpha,\bd\in\{1,2\}$.  The generators are
related by $(Q_\alpha)^\dagger = \bar{Q}_\ad$ and it is simply a
matter of notation that $Q$ is written with normal (undotted) indices
whereas $\bar{Q}$ is written with dotted indices.

If we allow for one set of such fermionic generators (corresponding to
$N=1$ supersymmetry) according to the {\it Haag-Lopuszanski-Sohnius
  theorem}~\cite{Haag:1974qh} we can in fact extend the Poincar\'e
algebra of Eqs.~(\ref{algPP})--(\ref{algMM}) in a non-trivial way to
the $N=1$ super Poincar\'e algebra:
\begin{eqnarray}
\left[Q_\alpha,P^\rho\right] &=& 0 \label{algQP} 
\\ \left\{Q_\alpha,\bar{Q}_\bd\right\} &=& 
2 (\sigma^\rho)_{\alpha\bd} P_\rho \label{algQQ} \\ 
\left[M^{\rho\sigma},Q_\alpha\right] &=& -i
(\sigma^{\rho\sigma})_\alpha^{\ \beta} Q_\beta
\label{MQ} \\
\left\{Q_\alpha,Q_\beta\right\} &=&
 \left\{\bar{Q}_\ad,\bar{Q}_\bd\right\} = 0
\label{algQQ0}
\end{eqnarray}
We could add another set of fermionic operators, ending up with $N=2$
supersymmetry, or in fact add even more sets. We will restrict
ourselves to $N=1$ however, because $N>1$ theories are ruled out as a
``low-energy'' (i.e. TeV) extension of the Standard Model, as will be
explained in Section~\ref{sec:ps}.

Note that the relations between two fermionic generators are given by
anticommutators, whereas relations involving at least one bosonic
operator involve the commutator. We will not delve into the derivation
of Eqs.~(\ref{algQP})--(\ref{algQQ0}). We only note that the addition
of fermionic generators also implies that we will have to increase the
set of coordinates (as we had to when extending
Eqs.~(\ref{PPalg})--(\ref{PJalg})), a point we will come back to in
Section~\ref{sec:superspace}.

It is important to realize what a strong motivation this provides. We
know that symmetries play a crucial role in physics and, in
particular, that the Poincar\'e symmetry is realized in nature. At the
same time, the only way to increase the Poincar\'e symmetry is
supersymmetry. It is for this reason that supersymmetry takes a
somewhat special status in the many possible scenarios of physics
beyond the Standard Model. We also remark that many motivations
usually mentioned, in particular the solution to the hierarchy
problem, are simply consequences of the increased symmetry in the
theory. While other approaches might solve the hierarchy problem as
well, susy was not initially introduced to solve this problem (nor to
unify gauge couplings).

\section{Weyl spinors and Grassmann variables \label{sec:tech}}

In this section we present the minimal amount of technicalities
required to be able to construct and write down supersymmetric and
Lorentz invariant theories in an efficient way. More details on the
conventions and notations used are given in
Appendix~\ref{App:Conventions}.

When dealing with fermions, we usually use Dirac spinors $\Psi(x)$
with four components. However, in susy theories it is more convenient
to work with Weyl spinors, $\psi(x)$ and $\chi(x)$, each with two
components only, writing
\begin{equation}
\Psi = \left(\begin{array}{c} \psi_\alpha
  \\ \chib^\ad \end{array}\right) ; \qquad
\overline{\Psi}  = 
\left(\chi^\alpha\ \ \psib_\ad \right);
\label{DiracToWeyl}
\end{equation}
Note that the bar over a Dirac spinor and a Weyl spinor mean something
different. For the Dirac spinor $\overline{\Psi}\equiv \Psi^\dagger
\gamma^0$ denotes the usual Dirac adjoint, whereas for Weyl spinors
the bar indicates that if $\psi_\alpha$ transforms with a certain
matrix $M$ under LT, $\psib_\ad$ transforms with the complex conjugate
matrix $M^*$, see \Eqn{app:WeylM}. Using the explicit form of
$\gamma^0$, \Eqn{app:gamma}, in \Eqn{DiracToWeyl} we find the precise
relation between them
\begin{equation}
\psib_\ad = \big[\psi_\alpha\big]^\dagger; \qquad
\chi^\alpha = \big[ \chib^\ad\big]^\dagger;
\label{WeylConjugation}
\end{equation}
The indices $\alpha$, $\ad$ run from 1 to 2 and, as for the generators
$Q$, it is simply a matter of notation that Weyl spinors corresponding
to the first two (last two) components of a Dirac spinor are written
with undotted (dotted) indices.

The helicity projection operators acting on a Dirac spinor yield
\begin{equation}
P_L \Psi \equiv \frac{1}{2}(1-\gamma_5)\Psi = \psi_\alpha\,; \quad
P_R \Psi \equiv \frac{1}{2}(1+\gamma_5)\Psi = \chib^\ad\,;
\label{leftrightproject}
\end{equation}
Thus, $\psi_\alpha$ and $\chib^\ad$ are called left-handed and
right-handed Weyl spinors respectively.  The indices of Weyl spinors
can be raised/lowered with the totally antisymmetric $\epsilon$-tensor,
\Eqn{app:raiselower}. The whole machinery is set up such that products
of Weyl spinors such as
\begin{eqnarray}
\chi\psi &\equiv& \chi^\alpha \psi_\alpha = \chi^\alpha
\epsilon_{\alpha\beta}  \psi^\beta 
\label{psiprod}\\
\chib \psib &\equiv& \chib_\ad \psib^\ad = \chib_\ad
\epsilon^{\ad\bd}  \psib_\bd
\label{psiBprod}
\end{eqnarray}
are Lorentz invariant. Note the different positions of the dotted and
undotted indices in the definition of the products.

Having written Dirac 4-spinors in terms of Weyl 2-spinors we have to
do the same for Dirac $4\times 4$ matrices. They are written in terms
of Pauli $2\times 2$ matrices $\sigma^\mu$ and the related matrices
$\sigmab^\mu$. The details are given in \Eqns{app:pauli}{app:gamma}.
What is important for us is that with this setup we are now able to
write the bilinear covariants that appear in Lagrangians in terms of
Weyl spinors. In particular we have
\begin{eqnarray}
\ol{\Psi} \Psi &=& \chi \psi + \psib \chib \equiv
  \chi^\alpha \psi_\alpha + \psib_\ad \chib^\ad \nn \\
\ol{\Psi} \gamma^\mu \Psi &=& \chi \sigma^\mu \chib 
- \psi \sigma^\mu \psib \equiv
  \chi^\alpha (\sigma^\mu)_{\alpha \ad} \chib^\ad - 
  \psi^\alpha (\sigma^\mu)_{\alpha \ad} \psib^\ad  
\end{eqnarray}
with a more complete list of relations given in \Eqn{app:covariants}.
Thus the standard Lagrangian for a free Dirac spinor can be written in
terms of Weyl spinors as
\begin{equation}
i\, \ol{\Psi} \gamma^\mu  \partial_\mu\Psi
-m\, \ol{\Psi} \Psi = 
i\, \chi \sigma^\mu \partial_\mu\chib 
+i\, \psi\sigma^\mu \partial_\mu\psib
-m\, \chi \psi -m\, \psib \chib
\label{diracWeyl}
\end{equation}
where we used integration by parts $-i\, (\partial_\mu\psi)\sigma^\mu
\psib = i\, \psi\sigma^\mu \partial_\mu\psib$. Sometimes identities
like $\psi \sigma^\mu \psib = - \psib \sigmab^\mu \psi$ are used to
write the kinetic part of the Lagrangian such that the r.h.s. of
\Eqn{diracWeyl} resembles more closely the l.h.s.

A Majorana spinor can be written in terms of a single Weyl spinor as 
\begin{equation}
\Psi_M = \left(\begin{array}{c} \psi_\alpha
  \\ \psib^\ad \end{array}\right) ; \qquad
\overline{\Psi}_M  = 
\left(\psi^\alpha\ \ \psib_\ad \right);
\label{MajoranaToWeyl}
\end{equation}
and the standard Lagrangian written in terms of  Weyl spinor reads
\begin{equation}
\frac{i}{2} \ol{\Psi}_M \gamma^\mu  \partial_\mu\Psi_M
-\frac{m}{2}\, \ol{\Psi}_M \Psi_M = 
\frac{i}{2}\Big(\psi\sigma^\mu\partial_\mu\psib - 
(\partial_\mu\psi)\sigma^\mu\psib\Big)
- \frac{m}{2} \Big(\psi\psi + \psib\psib\Big)
\label{majoranaWeyl}
\end{equation}
Of course, we could use integration by parts again, but prefer to
write the Lagrangian in symmetric form.

It might seem that we have made a step backwards in introducing Weyl
spinors, since the l.h.s. of the above equations clearly are more
compact than the r.h.s. However, the theories we are interested in
(i.e. supersymmetric extensions of the Standard Model) are
intrinsically chiral and it will turn out to be an advantage if this
is reflected in our formalism from the beginning. What is important to
realize is that expressions that look rather complicated, actually have
a very simple behaviour under Lorentz transformations. If all spinor
and all Lorentz indices are contracted, the expression is invariant
under Lorentz transformations. If there is one free Lorentz index, it
transforms as a four vector etc. Thus, simply by looking at the
expression we will be able to determine the transformation
property. This is an invaluable tool for constructing Lorentz invariant
Lagrangians and we want to have a similar formalism for constructing
supersymmetric Lagrangians.

In order to achieve this we have to introduce another technical tool,
Grassmann variables, or more precisely, Grassmann spinors. A Grassmann
variable (or fermionic variable) is like any other variable, except
that it anticommutes with other Grassmann variables (and commutes with
ordinary variables). This behaviour is similar to the behaviour of the
generators in the Poincare algebra Eqs.~(\ref{algQP})--(\ref{algQQ0}).
We can think of Grassmann variables as anticommuting complex numbers. 

A Grassmann spinor $\theta^\alpha$ or $\thb^\ad$ is made of two
Grassmann variables
\begin{equation}
\theta^\alpha = \left( \begin{array}{c} \theta^1 \\ \theta^2
\end{array} \right); \ \  
\thb^\ad = \left( \begin{array}{c} \thb^1 \\ \thb^2
\end{array} \right);
\label{GSdef}
\end{equation}
with each entry being a Grassmann variable,
i.e. $\{\theta^\alpha, \theta^\beta\}
= \{\theta^\alpha, \thb^\bd\} =  0$ and, in
particular $\theta^\alpha \theta^\alpha = 0$ ($\alpha\in\{1,2\}$, no
summation).  Note that in agreement with \Eqn{psiprod} the product of
a Grassmann spinor with itself is given by $\theta\theta =
\theta^1\theta_1+\theta^2\theta_2 = -2 \theta^1\theta^2$ and does not
vanish. However, adding one more factor of $\theta^\alpha$ does give
zero. This means that if we Taylor expand an arbitrary function
$\phi(\theta)$ in $\theta$ and include all terms up to the
$\theta\theta$ term, we actually reproduce the full function. Thus we
can parameterize any function $\phi(\theta)$ in terms of two
constants $c$ and $f$ and a constant Grassmann spinor $\zeta$ and
write
\begin{equation}
\phi(\theta) = c + \theta\zeta + f\, \theta\theta
\label{taylor}
\end{equation}
This will be important later on. 

We also remark that with the help of Grassmann spinors we can write
the super Poincar\'e algebra entirely in terms of commutators. In
particular we have in place of \Eqn{algQQ}
\begin{equation}
\big[\theta Q, \thb \bar Q\big] \equiv 
\big[\theta^\alpha Q_\alpha, \thb_\ad \bar Q^\ad\big] = 
2\, \theta\sigma^\mu\thb\, P_\mu
\label{QQcomm}
\end{equation}

Finally, we also need to introduce differentiation and integration
with respect to Grassmann variables. Derivatives with respect to
Grassmann variables are defined in \Eqn{app:gmder} and differentiating
e.g. $\phi(\theta)$ as given in \Eqn{taylor} with respect to
$\theta^\alpha$ we get $\partial_\alpha \phi \equiv
\partial/\partial\theta^\alpha\, \phi = \zeta_\alpha + 2 f
\theta_\alpha$. The integration is defined such that it always picks
out the highest part in the Taylor expansion of the function.  The
details are given in \Eqn{app:integrate}, but the only important fact
is that
\begin{eqnarray}
\int \phi(\theta) \, {\rm d}^2\theta &=&
\big[\phi(\theta)\big]_{\theta\theta} = f 
\label{intf} \\
\int \Omega(\theta,\thb) \, {\rm d}^2\theta {\rm d}^2\thb &=& 
\big[\Omega(\theta,\thb)\big]_{\theta\theta\, \thb\thb} =d 
\label{intd}
\end{eqnarray}
with $\phi(\theta)$ as given in \Eqn{taylor} and $d$ is the term
proportional to $\theta\theta\, \thb\thb$ in the double expansion of
the arbitrary function $\Omega(\theta,\thb)$ in $\theta$ and
$\thb$. We will actually never use the notation with the integral sign
and simply think of the operation $[\ldots ]_{\theta\theta}$ as
selecting the $\theta\theta$ component of the argument. It is not a
coincidence that the constants in \Eqns{intf}{intd} are denoted by $f$
and $d$ since -- as we will see later -- this is related to the common
terminology of $F$-terms and $D$-terms.

\section{Superspace and superfields \label{sec:superspace}}

Our starting point was to consider Poincar\'e symmetries. More
precisely, we write a Lagrangian as a function of fields $\phi(x)$
which have certain transformation properties under Poincar\'e
transformations, \Eqn{Poincare}. We then insist that the Lagrangian is
invariant under such transformations.

We also decided to enlarge our symmetry group with fermionic
generators. It is clear that in this case we also need some fermionic
coordinates that change in a certain way under the enlarged group of
transformations. Because we added the generators $Q_\alpha$ and $\bar
Q_\ad$ we will need a matching set of coordinates which we denote by
$\theta^\alpha$ and $\thb^\ad$. As a consequence, our fields will now
not only depend on $x^\mu$ but also on $\theta^\alpha$ and
$\thb^\ad$. We will write a generic field as $\Omega(x,\theta,\thb)$.
Such a field is called a {\it superfield} and the enlarged space is
called {\it superspace} with coordinates
$X=(x^\mu,\theta^\alpha,\thb^\ad)$. This extension of coordinates is
similar to the extension from $\vec{x}$ to $x^\mu = (t,\vec{x})$ in
Section~\ref{sec:sym}. Note that the mass dimension of the Grassmann
coordinates $\theta$ and $\thb$ is given by $[\theta] = [\thb] = -1/2$
whereas obviously $[x]=-1$.

Our ultimate goal is to construct Lagrangians that are invariant under
susy transformations. Thus we will need to get a handle on
the transformation property of fields. As a first step, we would like
to find a representation of the generators in terms of differentiation
operators, i.e. equations for $Q_\alpha$ and $\bar Q_\bd$ that are
analogous to $P_\mu = i\partial_\mu$.

Let us consider a susy transformation with $\omega^{\mu\nu}$ of
\Eqn{LT} set to zero for simplicity
\begin{equation}
S(a,\zeta,\zeb) \equiv 
e^{i\left(\zeta^\alpha Q_\alpha + \zeb_\ad \bar Q^\ad + a^\mu P_\mu\right)}
\label{susytrsf}
\end{equation}
with parameters $a$, $\zeta$ and $\zeb$ and where $Q$, $\bar Q$ and
$P$ are operators in Fock space. Note that if we set $\zeta=\zeb=0$
the transformation is simply a translation under which a quantum field
transforms as
\begin{equation}
\phi(x) \to S(a,0,0)\phi(x) S^{-1}(a,0,0) 
= e^{i a^\mu P_\mu}\phi(x) e^{-i a^\mu P_\mu}
= \phi(x+a)
\label{qftrsf}
\end{equation}
If we combine two susy transformations, we obtain
\begin{equation}
S(a,\zeta,\zeb)S(x,\theta,\thb) = 
S(x^\mu+a^\mu+i\, \zeta\sigma^\mu\thb -i\, \theta\sigma^\mu\zeb, \,
   \theta+\zeta,\, \thb+\zeb)
\label{combine_s}
\end{equation}
This can be derived by using the Baker-Campbell-Hausdorff formula
which states that if the commutators $[A,[A,B]]$ etc. vanish we have
$e^A e^B = e^{A+B+[A,B]/2}$. The only non-vanishing commutators we have
in deriving \Eqn{combine_s} are $[\zeta Q,\thb \bar Q] = 2\,
\zeta\sigma^\mu\thb\, P_\mu$ and $[\zeb \bar Q,\theta  Q] = -2\,
\theta\sigma^\mu\zeb\, P_\mu$. \Eqn{combine_s} states that even if we
set $a^\mu=x^\mu=0$ we induce a translation. This is a direct
consequence of \Eqn{algQQ}. Thus, starting from a point
$X=(x^\mu,\theta^\alpha,\thb^\ad)$ in superspace, under a susy
transformation, \Eqn{susytrsf} we have
\begin{equation}
X \to X' = 
(x^\mu+a^\mu+i\, \zeta\sigma^\mu\thb - i\,\theta\sigma^\mu\zeb, \,
\theta+\zeta,\, \thb+\zeb)
\label{Xtrsf}
\end{equation}
This is the generalization of \Eqn{Poincare}.

We now consider a superfield $\Omega(x,\theta,\thb)$ under a susy
transformation \Eqn{susytrsf}
\begin{eqnarray}
\hspace*{-1cm}
\Omega(x,\theta,\thb) &\to& 
e^{i\left(\zeta^\alpha Q_\alpha + \zeb_\ad \bar Q^\ad + a^\mu P_\mu\right)}
\, \Omega(x,\theta,\thb)\,
e^{-i\left(\zeta^\alpha Q_\alpha + \zeb_\ad \bar Q^\ad + a^\mu P_\mu\right)}
\nonumber \\
&& =\
\Omega(x^\mu+a^\mu +i\, \zeta\sigma^\mu\thb - i\, \theta\sigma^\mu\zeb, \,
\theta+\zeta,\, \thb+\zeb)
\label{Omegasusy}
\end{eqnarray}
Since we will need to calculate the transformation of fields several
times, we want to find a simple representation for \Eqn{Omegasusy}. We
seek differential operators $Q$, $\bar Q$ and $P$ such that the
transformation given in \Eqn{Omegasusy} can be written as
\begin{equation}
\Omega(x^\mu+a^\mu +i\, \zeta\sigma^\mu\thb - 
i\, \theta\sigma^\mu\zeb,\, \theta+\zeta,\, \thb+\zeb) = 
e^{-i\left(\zeta^\alpha Q_\alpha + \zeb_\ad \bar Q^\ad + a^\mu P_\mu\right)} 
\Omega(x,\theta,\thb)
\label{OmegaClassical}
\end{equation}
Note that this is quite some abuse of notation. In
\Eqn{OmegaClassical} $Q$, $\bar Q$ and $P$ are differential operators
that act on a function $\Omega(x,\theta,\thb)$, whereas in
\Eqns{Omegasusy}{qftrsf} $Q$, $\bar Q$ and $P$ are operators in Fock
space (i.e. can be written in terms of creation and annihilation
operators) and $\Omega$ is a quantum field, i.e. also an operator in
Fock space. It is customary but somewhat unfortunate to use the same
symbols for these different objects. Note that as far as $P$ is
concerned, \Eqn{OmegaClassical} is in agreement with
\Eqn{translation}. Indeed, we can combine $\Phi'(x') = \Phi(x)$ with
\Eqn{translation} to obtain $\Phi(x+a)=e^{-i\,a^\rho P_\rho}\,
\Phi(x)$. But we could change the sign and/or $i$ factors in the
coefficients multiplying $Q$ and $\bar Q$. This simply would lead to
different conventions for $Q$ and $\bar Q$ and, unfortunately, many
different conventions are used in the literature.

If we assume $a$, $\zeta$ and $\zeb$ to be infinitesimally
small we can Taylor expand both sides of \Eqn{OmegaClassical} (see
\Eqn{app:taylor}) 
\begin{eqnarray}
&& \Omega + 
\left(a^\mu + i\, \zeta\sigma^\mu\thb -i\, \theta\sigma^\mu\zeb \right)
\partial_\mu \Omega
+\zeta^\alpha \partial_\alpha\Omega
-\zeb_\ad \pab^\ad\Omega \nn \\
&& =\ \Omega -i\left(
\zeta^\alpha Q_\alpha + \zeb_\ad \bar Q^\ad + a^\mu P_\mu\right) \Omega
\label{Omegatrsf}
\end{eqnarray}
where $\Omega = \Omega(x,\theta,\thb)$. By comparing the coefficients
of the infinitesimal parameters $a^\mu$, $\zeta^\alpha$ and $\zeb^\ad$
we finally obtain
\begin{eqnarray}
P_\mu &=& i \partial_\mu \label{Prep} \\
Q_\alpha &=& i\, \partial_\alpha -
\sigma^\mu_{\alpha\ad} \thb^\ad\, \partial_\mu  \label{Qrep} \\
\bar{Q}_\ad &=& - i\, \pab_\ad + 
\theta^\alpha \sigma^\mu_{\alpha\ad}\,  \partial_\mu
\label{QBrep}
\end{eqnarray}
It is a useful exercise to check that these representations indeed
satisfy \Eqns{algQQ}{algQQ0}.  We can now use these expressions to
compute the change of a superfield $\Omega$ under a susy
transformation
\begin{equation}
\Omega\to \Omega' = \Omega+\delta \Omega = \Omega -i\left(
\zeta^\alpha Q_\alpha + \zeb_\ad \bar Q^\ad + a^\mu P_\mu\right) \Omega
\label{deltaOmega}
\end{equation}
For future reference we also introduce {\it covariant derivatives}
\begin{equation}
D_\alpha \equiv \partial_\alpha - i\,\sigma^\mu_{\alpha\ad} \thb^\ad
\, \partial_\mu\, ; \quad
\bar{D}_\ad \equiv \pab_\ad - i\,\theta^\alpha \sigma^\mu_{\alpha\ad}
\, \partial_\mu\, ;
\label{Ddef}
\end{equation}
defined such that they satisfy $\{D_\alpha, Q_\beta\} = \{D_\alpha,
\bar Q_\bd\} = 0$, with more relations given in \Eqn{app:algQD}. They
get their name from the fact that $D_\alpha\, \Omega$ (and $\bar
D_\ad\, \Omega$) transform in the same way\footnote{This is
  reminiscent of gauge theories, where the (gauge) covariant
  derivative $D_\mu$ is constructed such that a gauge field $\psi$ and
  $D_\mu \psi$ transform in the same way under gauge transformations.}
under susy transformation as $\Omega$, i.e. $D_\alpha\Omega\to
(D_\alpha\Omega)' = D_\alpha\Omega+\delta (D_\alpha\Omega)$ with
\begin{equation}
D_\alpha \delta\Omega =\delta (D_\alpha\Omega) =
 -i\left( \zeta^\alpha Q_\alpha + \zeb_\ad \bar Q^\ad 
+ a^\mu P_\mu\right) D_\alpha\Omega
\label{DOmegatrsf}
\end{equation}
We should warn the reader again that there are many different
conventions used in the literature and the explicit form of the
generators $Q_\alpha$ and $\bar{Q}_\ad$ and the covariant derivatives
$D_\alpha$ and $\bar D_\ad$ is by no means unique.

Let us now expand the most general superfield $\Omega(x,\theta,\thb)$
in $\theta$ and $\thb$. According to \Eqn{taylor} we expect terms with
one or two $\theta$ and/or $\thb$, but not more. Thus we write
\begin{eqnarray}
\Omega(x,\theta,\thb) &=& c(x) + \theta\psi(x) + \thb\psib'(x)
+(\theta\theta)\, F(x)  +  (\thb\thb)\, F'(x)
+\theta\sigma^\mu\thb\, v_\mu(x) \nn \\
&+&  (\theta\theta)\, \thb\lambdab'(x)
 + (\thb\thb)\, \theta\lambda(x)
+  (\theta\theta)\,(\thb\thb)\, D(x)
\label{Omega}
\end{eqnarray}
There are several points to be noted. First, the primed fields
e.g. $F'(x)$ are not in any way related to the corresponding unprimed
fields $F(x)$. They are simply the coefficients in the (terminating)
Taylor expansion of $\Omega$ in $\theta$ and $\thb$. Furthermore, it
is clear that there are four coefficients of the mixed
$\theta^\alpha\, \thb^\ad$ term. These four coefficients can
conveniently be written in terms of a vector field $v^\mu(x)$. Hence,
the superfield $\Omega$ contains four Weyl spinors $\psi$, $\psib'$,
$\lambda$ and $\lambdab'$, four scalar fields $c$, $F$, $F'$ and $D$
and a vector field $v$. These fields are called {\it component
  fields}.  Because a superfield contains a collection of component
fields it is often called a {\it supermultiplet}. There are eight complex
fermionic and eight complex bosonic degrees of freedom in $\Omega$. It
is of course not a coincidence that the number of bosonic and
fermionic degrees of freedom match.

The superfield $\Omega$ given in \Eqn{Omega} will not be one of the
basic blocks that we are going to use to construct supersymmetric
theories. We can define simpler building blocks by imposing
constraints. This will result in superfields with smaller particle
content. In the following two subsections we consider the two
important special cases.

\subsection{Chiral superfields} \label{sec:ChiralSF}

A superfield $\phi(x,\theta,\thb)$ that satisfies the constraint
$\bar{D}_\ad\, \phi(x,\theta,\thb) = 0$, where $\bar D$ is the
covariant derivative defined in \Eqn{Ddef}, is called a {\it
  left-handed chiral superfield} (LH$\chi$SF). The reason for the name
will become clear in a moment. Note that this constraint is self
consistent in the sense that it is invariant under susy
transformations. Indeed, after a susy transformation,
\Eqn{deltaOmega}, the superfield still satisfies the constraint. This
can be seen using \Eqn{DOmegatrsf}.

The constraint imposed reduces the number of degrees of freedom in the
superfield. To find the general expression of a LH$\chi$SF, analogous
to \Eqn{Omega}, we note that $\bar{D}_\ad \theta^\alpha = 0$ and
$\bar{D}_\ad\, y^\mu = 0$, where we define $y^\mu\equiv x^\mu
-i\,\theta\sigma^\mu\thb$. Thus, the most general function
$\phi(y,\theta) \equiv \phi(y) + \sqrt{2}\, \theta\psi(y) -
\theta\theta\, F(y)$ (the $\sqrt{2}$ and the minus sign are simply
conventions) satisfies $\bar{D}_\ad\, \phi = 0$. Expanding this back
in $x$, $\theta$ and $\thb$ we obtain
\begin{eqnarray}
\phi(x,\theta,\thb) &=& \varphi(x) + \sqrt{2} \theta \psi(x) -
i\theta\sigma^\mu\thb\, \partial_\mu \varphi(x) +\frac{i}{\sqrt{2}}
(\theta\theta) (\partial_\mu \psi(x)\sigma^\mu\thb) \nn \\
&-&\frac{1}{4} (\theta\theta)(\thb\thb) \partial^\mu\partial_\mu
\varphi(x) - (\theta\theta) F(x) 
\label{lhxsf}
\end{eqnarray}
as the expansion of a LH$\chi$SF into component fields. Again, we have
the same number of fermionic and bosonic degrees of freedom, with two
scalar fields $\varphi$ and $F$ and a Weyl spinor $\psi$. It is the
left-handed Weyl spinor $\psi$ that lends its name to the whole
superfield. The spinors of a LH$\chi$SF will be the left-handed quarks
and leptons of a susy extension of the Standard Model and the
$\varphi$ fields their supersymmetric partners, the squarks and
sleptons. The Higgs bosons and their susy partners will also form
chiral superfields. The mass dimension of the various component fields
in \Eqn{lhxsf} are $[\varphi]=1$, $[\psi]=3/2$ and $[F]=2$ such that
all terms in $\phi$ have mass dimension 1, i.e. $[\phi]=1$. Thus,
$\varphi$ and $\psi$ have the expected mass dimension, but $F$ does
not have the usual mass dimension of a scalar field. This is a first
hint that the $F$ component field is unphysical, an issue we will come
back to.

The susy transformation of a superfield, \Eqn{deltaOmega} induces
transformations of the component fields $\varphi(x) \to \varphi(x) +
\delta\varphi(x)$ etc. Using the explicit representation of $Q$ and
$\bar Q$, \Eqns{Qrep}{QBrep}, we find
\begin{eqnarray}
\delta\varphi &=& \sqrt{2}\, \zeta\psi \nn \\
\delta\psi_\alpha &=& - \sqrt{2}\, F\, \zeta_\alpha 
- i\sqrt{2}\, \sigma^\mu_{\alpha\ad}\zeb^\ad\, \partial_\mu\varphi
\label{deltacomponent} \\
\delta F &=& - i \sqrt{2}\, \partial_\mu\psi \sigma^\mu \zeb
=\partial_\mu \left( - i \sqrt{2}\,\psi \sigma^\mu \zeb\right)
\nn
\end{eqnarray}
As expected, the change in the bosonic/fermionic component fields is
proportional to the fermionic/bosonic fields. The crucial point is
that $\delta F$ is a total derivative. This will be very important
when we construct susy Lagrangians.

We can repeat the whole procedure for right-handed chiral superfields
(RH$\chi$SF) $\phi^\dagger$, which by definition satisfy the
constraint $D_\alpha\, \phi^\dagger = 0$. In terms of component fields
they read
\begin{eqnarray}
\phi^\dagger(x,\theta,\thb) &=& \varphi^\dagger(x) + \sqrt{2} \thb \psib(x) +
i\theta\sigma^\mu\thb\, \partial_\mu \varphi^\dagger(x) -\frac{i}{\sqrt{2}}
(\thb\thb) (\theta\sigma^\mu\partial_\mu \psib(x)) \nn \\
&-&\frac{1}{4} (\theta\theta)(\thb\thb) \partial^\mu\partial_\mu
\varphi^\dagger(x) - (\thb\thb) F^\dagger(x) 
\label{rhxsf}
\end{eqnarray}
The hermitian conjugate of a LH$\chi$SF is a  RH$\chi$SF.

\subsection{Vector superfields} \label{sec:VectorSF}

The chiral superfields ($\chi$SF) introduced above do not have a
vector field as component field. Thus, in order to deal with
supersymmetric gauge theories, we will also need another superfield,
called a {\it vector superfield} $V(x,\theta,\thb)$, that contains a
spin 1 component field. Such a superfield is defined by the constraint
$V(x,\theta,\thb) = V^\dagger(x,\theta,\thb)$. Again, this constraint
is preserved under susy transformations.

The expansion of a vector superfield (VSF) in terms of component
fields can be obtained by looking at \Eqn{Omega} and enforcing
$V=V^\dagger$.
\begin{eqnarray}
V(x,\theta,\thb) &=& c(x) + i\,\theta\chi(x) -i\,\thb\chib(x) +
\theta\sigma^\mu\thb\, v_\mu(x) 
+ i\, (\theta\theta) N(x) 
 - i\,(\thb\thb) N^\dagger(x) \nn \\
&+& i\, (\theta\theta) \thb
\left(\lambdab(x)+\frac{i}{2}\partial_\mu\chi(x)\sigma^\mu\right) 
 - i\, (\thb\thb) \theta 
\left(\lambda(x)-\frac{i}{2}\sigma^\mu\partial_\mu\chib(x)\right)  \nn \\
&+&  \frac{1}{2}(\theta\theta)(\thb\thb) \left( D(x) -
\frac{1}{2}\partial^\mu\partial_\mu c(x) \right) 
\label{vsf}
\end{eqnarray}
Several remarks are in order. First, factors $i$ and some overall
signs in the above expansion are simply conventions. Second, the
component fields $c$, $D$ and $v$ are now real, but $N$ is
complex. Thus, through the constraint $V^\dagger = V$ the eight
complex degrees of freedom in \Eqn{Omega} are reduced to eight real
bosonic and fermionic degrees of freedom in $V$. Putting it in other
words, in \Eqn{Omega}, the coefficients of e.g. $\theta$ and $\thb$,
denoted by $\psi$ and $\psib'$ were not related. However, in \Eqn{vsf}
the corresponding coefficients, denoted by $\chi$ and $\chib$ have to
be the same, i.e. there is only one Weyl spinor associated with the
$\theta$ term. The same is true for the $(\thb\thb)\, \theta$
term. In \Eqn{Omega} we denoted the corresponding component field by
$\lambda$, whereas in \Eqn{vsf} we redefine $\lambda$ such that the
coefficient takes a slightly more complicated form. The same remark
applies to the $\theta\theta\, \thb\thb$ term. The reason for this
will become clear in Section~\ref{sec:QED} and is related to the fact
that $V$ as given in \Eqn{vsf} has more degrees of freedom than we
bargained for. Apart from the vector field $v_\mu$ that we wanted (and
that gives the whole superfield its name and will represent gauge
bosons in susy extensions of the Standard Model) we might expect some
fermions (gauginos). However, we got two fermions, $\chi$ and
$\lambda$ and a whole set of scalar fields.  A look at the mass
dimension of the various component fields, $[c]=0$, $[\chi]=1/2$,
$[v]=[N]=1$, $[\lambda]=3/2$ and $[D]=2$ reveals that only $v$ and
$\lambda$ have the expected mass dimensions. Indeed, all other
component fields will turn out to be unphysical.

As we have done for the LH$\chi$SF in \Eqn{deltacomponent}, we could
now determine the transformation properties of the component fields of
$V$. However, as most component fields are unphysical, we refrain from
doing this and restrict ourselves to the transformation of the $D(x)$
component field. Under \Eqn{deltaOmega}, we have $D\to D+\delta D$ with
\begin{equation}
\delta D = \zeta\sigma^\mu\partial_\mu\lambdab(x)
 + \partial_\mu\lambda(x) \sigma^\mu\zeb
= \partial_\mu \left(
\zeta\sigma^\mu\lambdab(x) + \lambda(x) \sigma^\mu\zeb \right)
\label{Dtrsf}
\end{equation}
As for the $F$ field of a chiral superfield, the change in the $D$
field of a VSF is a total derivative.

\subsection{From superfields to particles} \label{sec:ps}

Let us pause for a moment an recapitulate what we have done. In
increasing the symmetry from the Poincar\'e group to the
super-Poincar\'e group we also had to increase the coordinate space
from Minkowski space with coordinates $x^\mu$ to superspace with
coordinates $X=(x^\mu,\theta^\alpha,\thb^\ad)$. Thus, our fields now
depend on $X$, i.e. not only on $x^\mu$ but also on $\theta^\alpha$
and $\thb^\ad$. In ``normal'' particle physics, the fields (e.g. the
electron or photon field) depend only on $x^\mu$. These ``normal''
fields are now simply the components of the superfields. Thus, susy
forces us to put several ``normal'' fields together into a superfield.

The most general expression for such a superfield is given in
\Eqn{Omega}. However, such a superfield is not a basic building block
for our theory since it contains too many component fields. We have
identified the three basic superfields that we will need in the
construction of susy extensions of the Standard Model. These are the
LH$\chi$SF, the RH$\chi$SF and the VSF. It will turn out that
ultimately the particle (i.e. ``normal'' field) content of the
LH$\chi$SF will be a scalar $\varphi$ and a left-handed fermion $\psi$
only. The other degree of freedom, the $F$-field will turn out to be
unphysical and will be eliminated. Similarly, for the RH$\chi$SF the
particle content is given by a scalar $\varphi^\dagger$ and a
right-handed fermion $\bar\psi$. In the case of the VSF, the particle
content will consist of a vector boson $v^\mu$ and a Weyl spinor
$\lambda$ with is conjugate $\lambdab$. All other fields will turn out
to be unphysical and will be eliminated.

Thus if we want to construct for example a susy version of QED, we
have to promote the left-handed (right-handed) electron field into a
LH$\chi$SF (RH$\chi$SF), thereby automatically introducing the scalar
partners, the selectrons. The photon field is embedded in a VSF which
introduces the fermionic partner of the photon, the photino. In the
case of the Standard Model we have
\begin{eqnarray}
\mbox{left-handed fermions:}& &\begin{array}{lllll}
\psi_f &\in& \phi_f &=& (\varphi_f,\psi_f)\end{array} 
 \label{lhf} \\[5pt]
\mbox{right-handed fermions:} & &\begin{array}{lllll}
  \psib_f &\in& \phi_f^\dagger &=& (\varphi_f^\dagger ,\psib_f)\end{array} 
 \label{rhf} \\[5pt]
\mbox{Higgs boson(s):} & & \begin{array}{lllll}
 \varphi_h\! &\in& \phi_h &=& (\varphi_h ,\psi_h) \\[2pt]
 \varphi_h^\dagger\! &\in& \phi_h^\dagger &=& (\varphi_h^\dagger ,\psib_h) 
 \end{array} 
 \label{hsf} \\[5pt]
\mbox{gauge bosons:}& & \begin{array}{lllll}
  v^\mu&\in& V &=& (v^\mu,\lambda, \lambdab)\end{array} 
 \label{vf} 
\end{eqnarray}
Thus, the leptons and quarks ($\psi_f$ and $\psib_f$) will be part of
a $\chi$SF ($\phi_f$ and $\phi^\dagger_f$) and get their scalar
partners, the sleptons and squarks ($\varphi_f$ and
$\varphi_f^\dagger$). The gauge bosons ($v^\mu$) will become a part of
a VSF ($V$) and will get their fermionic partners, the gauginos
($\lambda$ and $\lambdab$).  Finally the Higgs boson(s) ($\varphi_h$
and $\varphi_h^\dagger$) will be the scalar part of a $\chi$SF
($\phi_h$ and $\phi^\dagger_h$) and get their fermionic partners, the
higgsinos ($\psi_h$ and $\psib_h$). This will determine to a large
extent the particle content of the theory.

What we do not know yet is how to obtain the interactions between the
various particles of our theory. We have to make sure that these
interactions are compatible with susy. It is here where the superfield
formalism is an invaluable help, as we will see in the following
section.

Following up from our discussion just after \Eqn{algQQ0}, we can now
also understand why $N>1$ susy theories cannot be used as direct
low-energy extensions of the Standard Model. The nice feature about
$N=1$ is that it keeps the left-handed and right-handed fermions in
separate superfields as given in \Eqns{lhf}{rhf}. This is essential
because these fields transform differently under $SU(2)$ gauge
transformations. For $N>1$ the supermultiplets are larger and combine
the left-handed and right-handed fermions. This is inconsistent with
the weak interactions. Of course it is still possible that at very
high energies we have a $N>1$ theory. But this theory would have to be
broken such that at energy scales of a few TeV we have a $N=1$ susy
theory.

\section{Supersymmetric Lagrangians} \label{sec:Lagrangians}

The key observation for the construction of susy theories is that the
F-term of a chiral superfield (i.e. the $\theta\theta$ component of a
LH$\chi$SF or the $\thb\thb$ component of a RH$\chi$SF) and the D-term
of a VSF (i.e. the $\theta\theta\,\thb\thb$ component) transform into
themselves plus a total derivative under susy transformations. If the
Lagrangian ${\cal L}$ changes by a total derivative, the action
$\int{\rm d}^4x\, {\cal L}$ does not change at all. Thus, if we write
a Lagrangian as
\begin{equation}
{\cal L} = {\cal L}_F + {\cal L}_D
\label{LFD}
\end{equation}
where ${\cal L}_F$ is made up of F-terms (of $\chi$SF) and
${\cal L}_D$ is made up of D-terms (of VSF) we are guaranteed that our
theory is invariant under susy transformations. We will use this in
the following sections to construct various susy theories.

\subsection{The Wess-Zumino Lagrangian} \label{sec:WZ}

The Wess-Zumino model is the simplest susy Lagrangian and contains
only chiral superfields. If we have two LH$\chi$SF, $\phi_i$ and
$\phi_j$, then the product $\phi_i\phi_j$ is again a LH$\chi$SF, because
$\bar D_\ad (\phi_i\phi_j)= (\bar D_\ad \phi_i)\phi_j + \phi_i(\bar
D_\ad\phi_j) = 0$. Of course, this can be extended to an arbitrary
product of LH$\chi$SF and an equivalent statement holds for
RH$\chi$SF. Thus we define the {\it superpotential}
\begin{equation}
W(\phi_i) \equiv   a_i\, \phi_i + \frac{1}{2} m_{ij}\,
  \phi_i \phi_j + \frac{1}{3!} y_{ijk}\, \phi_i \phi_j \phi_k
\label{superpotential}
\end{equation}
where the sum $\sum_{ijk}$ over all possible combinations of
LH$\chi$SF is understood and $a_i$, $m_{ij}$ and $y_{ijk}$ are
constants. Then we can write
\begin{equation}
{\cal L}_{F, {\rm WZ}} = \int {\rm d}^2\theta\, W(\phi_i) 
+ \int {\rm d}^2\thb\, W^\dagger(\phi^\dagger_i) \equiv
\Big[W(\phi_i) \Big]_{\theta\theta} +
\Big[W^\dagger(\phi^\dagger_i) \Big]_{\thb\thb}
\label{WZf}
\end{equation}
The factors $1/2$ and $1/3!$ in \Eqn{superpotential} could be absorbed
into $m_{ij}$ and $y_{ijk}$ but usually are left explicit to take into
account the symmetry of the terms. According to \Eqn{intf}, the
integration ${\rm d}^2\theta$ picks out the $\theta\theta$ component,
hence ${\cal L}_{F, {\rm WZ}}$ results in a susy theory. One might
think we could add more terms with products of more than three
$\chi$SF in the superpotential and still end up with a susy
theory. However, this would result in a non-renormalizable
theory. Indeed, the mass dimension of the various couplings are
$[a_i]=2$, $[m_{ij}]=1$ and $[y_{ijk}]=0$ to ensure $[{\cal L}_{F,
    {\rm WZ}}]=4$. Had we added a term $c_{ijkl}\, \phi_i \phi_j
\phi_k\phi_l$ in \Eqn{superpotential} we would have a coupling with
negative mass dimension $[c_{ijkl}]=-1$.

We stress that ${\cal L}_{F, {\rm WZ}}$ contains arbitrary products of
LH$\chi$SF and arbitrary products of RH$\chi$SF but no terms like
$\phi_i \phi_j^\dagger$. This is of utmost importance and is due to
the fact that the $\theta\theta$ component (or the $\thb\thb$
component) of a term like $\phi_i \phi_j^\dagger$ does not transform
into itself plus a total derivative and hence would break susy. In
other words, the superpotential has to be a holomorphic (or analytic)
function of the superfields, i.e. it depends only on $\phi_i$ but not
on $\phi^\dagger_i$.

The Lagrangian ${\cal L}_{F, {\rm WZ}}$ as given in \Eqn{WZf} contains
mass terms and Yukawa couplings of the component fields, but no
kinetic terms, i.e. no terms like $(\partial_\mu\varphi_i)
(\partial^\mu\varphi_i)^\dagger$. It is clear that such terms can only
come from combinations of $\phi_i\, \phi^\dagger_i$ which we
explicitly excluded from the superpotential. On the other hand it is
also clear that $\phi_i\, \phi^\dagger_i$ is a vector superfield since
$(\phi_i\, \phi^\dagger_i)^\dagger =\phi_i\, \phi^\dagger_i $. Thus we
can get a supersymmetric Lagrangian by taking the D-term of
$\phi_i\,\phi^\dagger_i$.  Such a term has mass dimension 4. Higher
products such as $(\phi_i\,\phi^\dagger_i)(\phi_j\,\phi^\dagger_j)$
would lead to non-renormalizable interactions. Thus we write
\begin{equation}
{\cal L}_{D, {\rm WZ}} = \int {\rm d}^2\theta{\rm d}^2\thb\, 
\phi_i\,\phi^\dagger_i
=  \left[\phi_i\,\phi^\dagger_i\right]_{\theta\theta\, \thb\thb}
\label{WZd}
\end{equation}
and the full Lagrangian ${\cal L}_{\rm WZ} ={\cal L}_{F, {\rm WZ}}
+{\cal L}_{D, {\rm WZ}} $ has the structure given in \Eqn{LFD}.

The usefulness of \Eqns{WZf}{WZd} lies in the fact that a simple
glance immediately reveals that the theory is supersymmetric. On the
other hand, \Eqns{WZf}{WZd} are fairly useless if we want information
about the particle content and interactions of the theory. To
obtain this we will have to express ${\cal L}_{{\rm WZ}}$ in terms of
component fields. Given the explicit expression \Eqns{lhxsf}{rhxsf}
this is trivial if slightly tedious (for details see
Appendix~\ref{app:sample}). Considering the simplest case with only
one chiral superfield (and $a_1=a$, $m_{11}=m$, $y_{111}=y$) we get
\begin{eqnarray}
{\cal L}_{D, {\rm WZ}} &=&  F^\dagger F +
(\partial_\mu\varphi)\, (\partial^\mu\varphi)^\dagger + 
\frac{i}{2}\, \psi\sigma^\mu(\partial_\mu\psib)  -
\frac{i}{2}\, (\partial_\mu\psi)\sigma^\mu\psib
\label{WZdcomp} \\
{\cal L}_{F, {\rm WZ}} &=& 
-a\, F - m\, \varphi F - \frac{m}{2} (\psi\psi)
-\frac{y}{2}\,\varphi  \varphi F
-\frac{y}{2}\,\varphi (\psi\psi)  + {\rm h.c.}
\label{WZfcomp}
\end{eqnarray}
As expected, the D-term contains the kinetic term of the $\varphi$ and
the $\psi$ component fields (see \Eqn{majoranaWeyl}). Note however,
that there is no kinetic term for the $F$ field. This means that the
equation of motion for $F$ (and $F^\dagger$) reduces to an algebraic
equation
\begin{equation}
0 = \partial_\mu
\frac{\partial{\cal L}}{\partial(\partial_\mu F)} - 
\frac{\partial{\cal L}}{\partial F} = 
- \frac{\partial{\cal L}}{\partial F} =
- F^\dagger + a + m\, \varphi + \frac{y}{2}\, \varphi\varphi
\label{Feom}
\end{equation}
We can solve this trivially and eliminate $F$ and $F^\dagger$ from the
Lagrangian. The terms containing $F$ and $F^\dagger$ in
\Eqns{WZdcomp}{WZfcomp} then read
\begin{equation}
 F^\dagger F-\left(a\, F + m\, \varphi\, F +\frac{y}{2}\,\varphi
 \varphi\, F + {\rm h.c.}\right) =
- \left|a + m\, \varphi +\frac{y}{2}\,\varphi \varphi\right|^2 = 
-\left|\frac{\partial W(\varphi)}{\partial \varphi}\right|^2
\label{Feliminated}
\end{equation}
In the last step, $W(\varphi)$ is the usual superpotential, but it is
considered to be a function of the scalar component field $\varphi$
only, rather than the full superfield $\phi$. For writing a Lagrangian
in terms of component fields, this is usually more useful.

Performing a shift $\varphi\to \varphi + (M-m)/y$ with $M\equiv
\sqrt{m^2-2 a\, y}$ to eliminate the $a\, \varphi$ term (or simply
setting $a=0$) the Lagrangian reads
\begin{eqnarray}
{\cal L}_{\rm WZ} &=& 
(\partial_\mu\varphi)\, (\partial^\mu\varphi)^\dagger + 
\frac{i}{2}\, \psi\sigma^\mu(\partial_\mu\psib)  -
\frac{i}{2}\, (\partial_\mu\psi)\sigma^\mu\psib \nn \\
&-& |M|^2 \varphi\varphi^\dagger 
-\frac{|y|^2}{4} \varphi\varphi\varphi^\dagger\varphi^\dagger
- \left(\frac{M}{2} \psi\psi 
+ \frac{M^* y}{2} \varphi\varphi\varphi^\dagger
+\frac{y}{2} \varphi\, \psi\psi
+ {\rm  h.c.} \right)
\label{WZfinal}
\end{eqnarray}
This theory contains a spin 0 and a spin 1/2 particle with the same
mass. There is a three-point and a four-point interaction between the
scalars and a scalar-scalar-fermion interaction. The couplings of
these interactions are all related. Of course, this is simply a
consequence of susy.

For future reference, let us rewrite the Lagrangian in yet another
way. We will do this for the general case with an arbitrary number of
chiral superfields. 
\begin{eqnarray}
{\cal L}_{\rm WZ} &=& 
(\partial_\mu\varphi_i)\, (\partial^\mu\varphi_i)^\dagger + 
\frac{i}{2}\, \psi_i\sigma^\mu(\partial_\mu\psib_i)  -
\frac{i}{2}\, (\partial_\mu\psi_i)\sigma^\mu\psib_i \nn \\
&-& \sum_i \left|\frac{\partial W(\varphi_i)}{\partial \varphi_i}\right|^2
- \frac{1}{2} \left(\frac{\partial^2 W(\varphi_i)}{\partial \varphi_i\,
  \partial \varphi_j}\right) \psi_i \psi_j 
- \frac{1}{2} \left(\frac{\partial^2 W^\dagger(\varphi_i)}{\partial
  \varphi^\dagger_i\,  \partial \varphi^\dagger_j}\right) \psib_i \psib_j 
\label{WZgen}
\end{eqnarray}
The superpotential is as given in \Eqn{superpotential} but considered
to be a function of the scalar component fields $\varphi_i$ only. Note
that the superpotential determines all interactions and the mass terms
of the component fields, and thus, the full theory.

\subsection{Susy QED} \label{sec:QED}

The Wess-Zumino Lagrangian does not contain spin 1 component
fields. Thus, to obtain susy gauge theories we will have to extend the
field content and include VSF. If we have a VSF $V=V^\dagger$ then
$V^n$ is also a VSF and its D-term (i.e. its $\theta\theta\, \thb\thb$
component) is supersymmetric. However, this will not lead to kinetic
terms for the corresponding spin~1 vector field $v^\mu$. As in the
case of chiral superfields we will have to add another construct for
the kinetic terms. We define\footnote{In the literature usually the
  notation $W_\alpha$ and $\bar W_\ad$ is used in \Eqn{Udef}. We use
  $U_\alpha$ and $\bar U_\ad$ to avoid confusion with the
  superpotential.}
\begin{equation}
U_\alpha \equiv -\frac{1}{4} \big(\bar D \bar D\big) D_\alpha V\, ;
\qquad
\bar U_\ad \equiv -\frac{1}{4} \big(D D\big) \bar D_\ad V\, ;
\label{Udef}
\end{equation}
Because of $\bar D_\ad \bar D \bar D = 0$ we know that $U_\alpha$ is a
LH$\chi$SF, $\bar D_\ad U_\alpha = 0$. Similarly, $\bar U_\ad$ is a
RH$\chi$SF. Forming the products $U^\alpha U_\alpha$ and $\bar U_\ad
\bar U^\ad$ as in \Eqns{psiprod}{psiBprod} we obtain a Lorentz
invariant expression. Furthermore, the corresponding F-terms are
supersymmetric and in fact they do contain the kinetic terms of the
component fields $v^\mu$ and $\lambda$ (see
Appendix~\ref{app:sample}).

Before we look at this in more detail we have to combine gauge
symmetry with susy. After all, our vector bosons are supposed to be
gauge bosons. Let us start with a global $U(1)$ gauge symmetry. Under
such a symmetry, component fields transform as $\varphi
\rightarrowtail\varphi' = e^{-i\Lambda}\varphi$ where $\Lambda$ is a
real constant and has mass dimension $[\Lambda]=0$. It follows that
$\varphi^\dagger\varphi$ is gauge independent. We can easily extend
this to superfields by noting that a real constant
$\Lambda=\Lambda^\dagger$ is a special case of a chiral superfield. In
fact it is actually a LH$\chi$SF and a RH$\chi$SF at the same time
because $\bar D_\ad \Lambda = D_\alpha \Lambda = 0$. Thus a LH$\chi$SF
transforms as $\phi \rightarrowtail \phi' = e^{-i\Lambda}\phi$ with
$\phi'$ still being a LH$\chi$SF and $\phi^\dagger$ transforms as
$\phi^\dagger \rightarrowtail \phi'^\dagger =
e^{i\Lambda^\dagger}\phi^\dagger$ with $\phi'^\dagger$ still being a
RH$\chi$SF and $\big[\phi^\dagger\phi\big]_{\theta\theta\, \thb\thb}$
is supersymmetric and invariant under global gauge transformations.

If we want local gauge invariance, then $\Lambda$ will have to be a
function of $x$. We still want $\Lambda(x)$ ($\Lambda^\dagger(x)$)
to be a LH$\chi$SF (RH$\chi$SF) such that $\phi'$ ($\phi'^\dagger$) is
a LH$\chi$SF (RH$\chi$SF). However, it is not possible to have a
$x$-dependent superfield that is at the same time a LH$\chi$SF and a
RH$\chi$SF, thus we  have $\Lambda(x) \neq \Lambda^\dagger(x)$. As a
consequence, under gauge transformations 
\begin{equation}
\phi^\dagger\,\phi \rightarrowtail
\phi'^\dagger\,\phi' = 
\phi^\dagger\, e^{i\Lambda^\dagger(x)}\,e^{-i\Lambda(x)}\,\phi
\neq \phi^\dagger\,\phi
\label{abeliangauge}
\end{equation}
This seems to introduce new particles, the component fields of
$\Lambda$. However, they have the ``wrong'' mass dimension. Because
$\Lambda$ appears in the exponent, we must have $[\Lambda]=0$. This
entails mass dimensions 0 and 1/2 for the scalar and fermion component
fields of the $\chi$SF $\Lambda$, in contrast to the usual dimensions
1 and 3/2. As we will see, these component fields are unphysical and
can be eliminated together with the unphysical component fields of $V$.

According to \Eqn{abeliangauge} $\phi^\dagger\phi$ is invariant under
global but not local gauge transformations. This is of course very
familiar from standard non-susy theories, where
e.g. $(\partial_\mu\varphi)^\dagger (\partial^\mu\varphi)$ is
invariant under global but not local gauge transformations. As in
these cases, to restore local gauge invariance we have to introduce a
gauge VSF, $V$, transforming under gauge transformations as
\begin{equation}
e^V \rightarrowtail e^{-i \Lambda^\dagger(x)}\, e^V\, e^{i \Lambda(x)}
\label{gaugetrsf}
\end{equation}
Note that in the abelian case, where all superfields commute, this can
be written as
\begin{equation}
V \rightarrowtail V' = V - i \Lambda^\dagger(x) +i \Lambda(x)
\label{abeliangtrsf}
\end{equation}
Then the term 
\begin{equation}
\big[ \phi^\dagger\, e^V\,
  \phi\big]_{\theta\theta\, \thb\thb}\rightarrowtail
\big[ \phi'^\dagger\, e^{V'}\,
  \phi'\big]_{\theta\theta\, \thb\thb} = 
\big[ \phi^\dagger\, e^V\,
  \phi\big]_{\theta\theta\, \thb\thb}
\label{kineticabelian}
\end{equation}
is supersymmetric and invariant under local gauge transformations.

The general expression of a VSF in terms of component fields is given
in \Eqn{vsf}. We can exploit the gauge transformation
\Eqn{abeliangtrsf} to obtain a particularly convenient representation
of the gauge VSF. If we choose $\Lambda(x,\theta,\thb)$ as in
\Eqn{lhxsf} but with the replacements $\psi\to -\chi/\sqrt{2}$, $F\to
N$ and ${\rm Im}(\varphi)\to c/2$ we get for $V'\equiv V_{\rm WZ}$ the
simple expression
\begin{equation}
V_{\rm WZ}(x,\theta,\thb) = \theta\sigma^\mu\thb\, v_\mu(x)
+ i (\theta\theta)\, \thb\lambdab(x)
 - i (\thb\thb)\, \theta\lambda(x)
+  \frac{1}{2}(\theta\theta)(\thb\thb)\,  D(x)
\label{WZgauge}
\end{equation}
Note that we can also eliminate one degree of freedom in $v_\mu$
through a choice of ${\rm Re}(\varphi)$. Thus, we are left with four
(three in $v_\mu$ one in $D$) real bosonic and four real fermionic
degrees of freedom in $V_{\rm WZ}(x,\theta,\thb)$. This gauge is
called the {\it Wess-Zumino gauge} and has the nice feature that many
unphysical component fields of $V$ (and $\Lambda$) are eliminated. In
this respect it is reminiscent of the unitary gauge. We should remark
however, that this gauge choice is not invariant under susy
transformations. Indeed, if we compute the change $\delta V_{\rm WZ} =
-i(\zeta Q + \zeb \bar Q) V_{\rm WZ}$ under an infinitesimal pure susy
transformation, among many others, a term like $-i\, \zeta Q\ (-i\,
\thb\thb\, \theta\lambda) = -i\, \theta\theta\, \zeta\lambda$ is
generated. Such a term corresponds to a $N$ component field in
\Eqn{vsf} which is not present in \Eqn{WZgauge}.

In order to complete the construction of an abelian supersymmetric
gauge theory, we note that $U_\alpha$ and $\bar U_\ad$ are gauge
independent. This can be verified by using \Eqn{abeliangtrsf} in
\Eqn{Udef} and using $\bar D_\ad \Lambda = D_\alpha \Lambda^\dagger =
0$ (see Appendix~\ref{app:sample}). Thus we have an abelian gauge
invariant and susy Lagrangian
\begin{equation}
{\cal L} =
\frac{1}{4}\big[ U^\alpha U_\alpha\big]_{\theta\theta} 
+ \frac{1}{4}\big[ \bar U_\ad \bar U^\ad\big]_{\thb\thb} 
+  \big[
\phi_i^\dagger\, e^{2 g\, V}\, \phi_i \big]_{\theta\theta\, \thb\thb}+ 
\big[W(\phi_i)\big]_{\theta\theta} 
+ \big[W^\dagger(\phi^\dagger_i)\big]_{\thb\thb}
\label{susyabelian}
\end{equation}
as long as we make sure the superpotential is gauge independent. In
particular, the fields present in the term $a_i\, \phi_i$ in
\Eqn{superpotential} have to be gauge singlets. In \Eqn{susyabelian}
$g$ denotes the gauge coupling and the normalization of the various
terms has been chosen such that we will recover the standard
normalization if we rewrite \Eqn{susyabelian} in terms of the
component fields.

If we consider QED, the $\chi$SF would correspond to a superfield for
each charged lepton. Thus we have a LH$\chi$SF, $\phi_1$, containing
the left-handed electron (as $\psi$) and its susy partner, the
``left-handed'' selectron (as $\varphi$). Note that the term
left-handed for the selectron is widely used but misleading, because
the spin of the selectron is 0. There is also the corresponding
RH$\chi$SF, $\phi^\dagger_1$, containing the right-handed electron (as
$\psib$) and its susy partner, the ``right-handed'' selectron (as
$\varphi^\dagger$). If we want to include the second and third family,
we have to introduce $\phi_2$ and $\phi_3$ as well as $\phi_2^\dagger$
and $\phi_3^\dagger$ containing the muons and taus respectively. In
this theory there cannot be a term $a_i\, \phi_i$ because none of the
fields is a gauge singlet. We could introduce one (or three)
LH$\chi$SF for the neutrino(s). Since they are singlets under
$U_{QED}(1)$, a linear term in the superpotential with these
LH$\chi$SF would be allowed. However, it is clear that introducing a
neutrino field in QED is not particularly interesting.

Let us consider the structure of the Lagrangian \Eqn{susyabelian} and
its form in terms of the component fields. The first two terms of
\Eqn{susyabelian} contain only the gauge boson $v^\mu$ (the photon),
its susy partner $\lambda$ (the photino) and the scalar $D$ field. As
we will see below, these terms are nothing but the kinetic terms of
the photon and photino.  The third term of \Eqn{susyabelian} can be
split into two parts. If we take the leading part of $e^{2g\, V} =
1+\ldots$, we see that this terms coincides with \Eqn{WZd} which in
component form is given in \Eqn{WZdcomp}. Thus it contains the kinetic
terms of the leptons and sleptons. The higher order terms in
$e^{2g\,V} = 2g\, V + \ldots$ contain the interactions between the
leptons (and sleptons) with the photon (and photino). Finally, the last
two terms of \Eqn{susyabelian} are again equivalent to the
corresponding terms discussed in Section~\ref{sec:WZ} and contain the
interactions involving only component fields of the $\chi$SF. In the
case of QED, the total charge of each term has to vanish to preserve
gauge invariance.

Let us consider the kinetic terms of the photon and photino in more
detail. The most tedious part of the calculation is to obtain an
expression for $U_\alpha$ in terms of the component fields. For this
(details are given in Appendix~\ref{app:sample}) it is convenient to
write $x^\mu$ in terms of $y^\mu$, as used in the derivation of
\Eqn{lhxsf} or $\bar y^\mu\equiv x^\mu +i\,\theta\sigma^\mu\thb$ which
satisfies $D_\alpha\, \bar y^\mu = 0$ and we obtain
\begin{equation}
U_\alpha = -i\, \lambda_\alpha(y) 
- \theta\theta\, \sigma^\nu_{\alpha\bd}\, \partial_\nu\bar\lambda^\bd(y)
- \frac{i}{2}\,\theta_\beta\, 
(\sigma^\mu\sigmab^\nu)_\alpha^{\ \beta} F_{\mu\nu}(y)
+ \theta_\alpha\, D(y)
\label{Uabelian}
\end{equation}
where $F_{\mu\nu} \equiv \partial_\mu v_\nu - \partial_\nu v_\mu$
is the usual field strength tensor and the component fields are
functions of $y^\mu= x^\mu - i\theta\sigma^\mu\thb$. Thus the first
two terms of \Eqn{susyabelian} in terms of the component fields are
given by
\begin{equation}
\frac{1}{4}\big[ U^\alpha U_\alpha\big]_{\theta\theta} 
+ \frac{1}{4}\big[ \bar U_\ad \bar U^\ad\big]_{\thb\thb}  = 
- \frac{1}{4} F^{\mu\nu}F_{\mu\nu} 
- \frac{i}{2} (\partial_\mu\lambda)\sigma^\mu\lambdab
+ \frac{i}{2} \lambda\sigma^\mu(\partial_\mu\lambdab)
+ \frac{1}{2} D^2
\label{FFabelian}
\end{equation}
and, indeed, contain kinetic terms for $v_\mu$ and $\lambda$.
However, there is no kinetic term for the $D$ component field. This
field is an auxiliary field, similar to the $F$ component field of
$\chi$SF, and will be eliminated using the equation of motion. Before
we can do this, we have to find all other terms containing $D$. They
are in the third term of \Eqn{susyabelian}. Note that in the
Wess-Zumino gauge $e^{2g\, V} = 1 + 2g\, V + 2 g^2\, V^2$, i.e. we
need at most two factors of $V$, because $V_{\rm WZ}^3$ and higher
powers vanish. We postpone the derivation of the full interaction term
to Section~\ref{sec:QCD} and write here only the term containing the
$D$ component field
\begin{equation}
\big[
\phi_i^\dagger\, e^{2g\, V}\, \phi_i \big]_{\theta\theta\, \thb\thb} = 
g\, \varphi_i^\dagger\, \varphi_i\, D + 
{\rm terms\ without\ } D
\label{abelianIA}
\end{equation}
In this context we mention that we can add another susy and gauge
invariant term to \Eqn{susyabelian}. We know already from \Eqn{Dtrsf}
that the $\theta\theta\, \thb\thb$ component of a VSF is susy. In the
case of an abelian gauge field, this term is also gauge
invariant. Indeed, \Eqn{abeliangtrsf} reveals that under a gauge
transformation the $\theta\theta\, \thb\thb$ component of a VSF
transforms into itself plus a total derivative, because the
$\theta\theta\, \thb\thb$ component of a $\chi$SF ($\Lambda$ and
$\Lambda^\dagger$ of \Eqn{abeliangtrsf}) are total derivatives. Thus
we could add a term
\begin{equation}
{\cal L}_{\rm FI} = 2\,\big[k\, V\big]_{\theta\theta\, \thb\thb}
= k\, D
\label{FIterm}
\end{equation}
to the Lagrangian \Eqn{susyabelian}, where $k$ is a constant (often
denoted by $\xi$ in the literature) with mass dimension $[k]=2$ and
the factor 2 is added for convenience.  Such a term is called a {\it
  Fayet-Iliopoulos term}~\cite{Fayet:1974jb} and will be important
later on when we discuss spontaneous breaking of susy. For the moment
we simply note that this term also depends on the component field $D$
as indicated in \Eqn{FIterm}.

The full Lagrangian ${\cal L} + {\cal L}_{\rm FI}$ does not contain
terms involving $\partial_\mu D$. Thus the equation of motion for $D$
is algebraic and can be solved trivially, resulting in
\begin{equation}
0= \frac{\partial {\cal L}}{\partial\, D} = 
\frac{\partial}{\partial\, D} \left(\frac{D^2}{2} +
g\,\varphi_i^\dagger \varphi_i\, D + k\, D\right) =
D+ g\, \varphi_i^\dagger \varphi_i + k
\label{eomD}
\end{equation}
As for the $F$ component field, we can solve this and eliminate the
$D$ component field from the Lagrangian. We obtain
\begin{equation}
\frac{D^2}{2} +  D \left(g\,\varphi_i^\dagger \varphi_i+k \right)
= -\frac{1}{2} \left(g\, \varphi_i^\dagger \varphi_i+k\right)^2
\label{Deliminated}
\end{equation}
for the terms containing the $D$ field in Eqs.~(\ref{FFabelian}),
(\ref{abelianIA}) and (\ref{FIterm}). This is analogous to
\Eqn{Feliminated}.

We refrain from writing down the full Lagrangian in terms of the
component fields. This will be done in the next section for a
non-abelian gauge theory from which the abelian limit can easily be
taken. 

\subsection{Susy QCD} \label{sec:QCD}

The construction of supersymmetric non-abelian gauge theories is
slightly more complicated, as expected. Without loss of generality we
will start by looking at $SU(3)$ with the eight generators $T^a$ and
the corresponding gauge superfields (containing the gluon) $V^a$. We
also introduce $V \equiv V^a T^a$ (where the sum $\sum_a$ with
$a\in\{1\ldots 8\}$ is understood) with the generators in the adjoint
representation and the gauge coupling $g$. The gauge transformation is
as given in \Eqn{gaugetrsf} with $\Lambda\equiv \Lambda^a T^a$. Note,
however, that \Eqn{abeliangtrsf} is not applicable any longer, due to
non-commuting terms in the Baker-Campbell-Hausdorff formula (see
remark after \Eqn{combine_s}).

We have to modify the kinetic terms, because $U_\alpha$ as defined in
\Eqn{Udef} is not gauge invariant in the non-abelian case. Instead we
define
\begin{equation}
U_\alpha \equiv 
-\frac{1}{8\, g} \bar D \bar D\,
e^{-2 g\, V} D_\alpha\, e^{2 g\, V}\, ;
\qquad
\bar U_\ad \equiv
\frac{1}{8\, g} D D\,
e^{2 g\,V} \bar D_\ad\, e^{-2 g\,V}\, ;
\label{Udefnonab}
\end{equation}
where again $U_\alpha \equiv U^a_\alpha\, T^a$ and $\bar U_\ad \equiv
\bar U^a_\ad\, T^a$. Using the expansion of the exponentials with
$V_{\rm WZ}^3 = V_{\rm WZ}^2 (D_\alpha V_{\rm WZ}) =0$ and $D_\alpha
V^2_{\rm WZ} =(D_\alpha V_{\rm WZ})V_{\rm WZ} +V_{\rm WZ}(D_\alpha
V_{\rm WZ})$ we can write \Eqn{Udefnonab} as
\begin{equation}
U_\alpha \equiv -\frac{1}{4}\bar D \bar D
\left(D_\alpha V + g [D_\alpha V,V]\right)\, ;
\qquad
\bar U_\ad \equiv -\frac{1}{4}D D
\left(\bar D_\ad V - g [\bar D_\ad V,V]\right)\, ;
\label{Unaexp}
\end{equation}
Thus, in the abelian case \Eqn{Udefnonab} reduces to \Eqn{Udef}, but
in the non-abelian case there is a difference due to $[T^a, T^b] \neq
0$, resulting in $[D_\alpha V,V]\neq 0$. Note that $U_\alpha$ and
$\bar U_\ad$ as given in \Eqn{Udefnonab} are not invariant under
non-abelian gauge transformations, but they transform like (see
Appendix~\ref{app:sample})
\begin{equation}
U_\alpha \rightarrowtail 
e^{-2 i g\, \Lambda}\,U_\alpha\, e^{2 i g\,  \Lambda}\, ; \qquad
\bar U_\ad \rightarrowtail 
e^{-2 i g\, \Lambda^\dagger}\,\bar U_\ad\, e^{2 i g\, \Lambda^\dagger}\, ;
\label{Utrsf}
\end{equation}
such that the trace (over the gauge group indices), ${\rm Tr}\,
U^\alpha U_\alpha = 1/2\, (U^a)^\alpha\, U^a_\alpha$ is gauge
invariant\footnote{We use the normalization ${\rm Tr}\, T^a T^b =
  \delta^{ab}/2$}. This is completely analogous to the non-susy case,
where the field-strength tensor $F_{\mu\nu}$ itself is invariant in
the abelian case, but in the non-abelian case only the trace ${\rm
  Tr}\, F^{\mu\nu} F_{\mu\nu} = 1/2\, (F^a)^{\mu\nu}\, F^a_{\mu\nu}$
is invariant, with $F_{\mu\nu}\equiv F^a_{\mu\nu} T^a$.

In the derivations above we have tacitly assumed that we can use the
Wess-Zumino gauge again. However, this is not clear a priori. After
all, \Eqn{abeliangtrsf} is not applicable in the non-abelian case. If we
use the Baker-Campbell-Hausdorff formula in \Eqn{gaugetrsf} we see
that the non-abelian generalization of \Eqn{abeliangtrsf} reads
\begin{equation}
V \rightarrowtail V' = V  +i (\Lambda-\Lambda^\dagger)
-\frac{i}{2}[\Lambda+\Lambda^\dagger,V]+ \ldots
\label{nabeliangtrsf}
\end{equation}
where we have left out an infinite tower of higher commutators
$[V,[V \ldots[V,(\Lambda-\Lambda^\dagger)]]$. Thus the relation
between $V$ and $\Lambda$ and  $\Lambda^\dagger$ in the Wess-Zumino
gauge fixing is more complicated, but we can still arrange $\Lambda$
and $\Lambda^\dagger$ such that $V'$ takes the form given in
\Eqn{WZgauge}.

It might not be obvious that $U_\alpha$ as defined in \Eqn{Udefnonab}
has the structure $U^a_\alpha\, T^a$. But the situation is again very
similar to $F^{\mu\nu}$. Performing an explicit computation (see
Appendix~\ref{app:sample}), we get terms involving commutators
$[T^a,T^b]$ which are written in terms of the structure constants,
using \Eqn{algTT}, and we get
\begin{equation}
U^a_\alpha = - \frac{i}{2}\,
 \theta_\beta (\sigma^\mu\sigmab^\nu)_\alpha^{\ \beta}\, F^a_{\mu\nu}
-\theta\theta\, \sigma^\mu_{\alpha\ad} (D_\mu \lambdab^a)^\ad
-i\, \lambda_\alpha^a + \theta_\alpha\, D^a
\label{Unonabelian}
\end{equation}
where the explicit form of the field-strength tensor and the (gauge)
covariant derivatives  are given by
\begin{eqnarray}
F^a_{\mu\nu} &\equiv& 
\partial_\mu v^a_\nu -  \partial_\nu v^a_\mu -
g f^{abc}\, v_\mu^b\, v_\nu^c  
\label{Fnonabelian}
\\
(D^\mu \lambdab^a)^\ad &\equiv&  
(\partial^\mu \lambdab^a)^\ad -
g f^{abc}\, (v^b)^{\mu} (\lambdab^c)^\ad
\label{covder}
\end{eqnarray}
and the component fields are functions of $y^\mu= x^\mu -
i\theta\sigma^\mu\thb$.  Note that the normalization and the details
of the definition in \Eqn{Udefnonab} have been chosen such that
\Eqn{Unonabelian} agrees with \Eqn{Uabelian} in the abelian limit
$f^{abc}\to 0$.

We can now proceed as in non-susy gauge theories and introduce an
arbitrary number of matter fields, in our case $\chi$SF, that
transform under a certain representation
\begin{equation}
\phi_i \rightarrowtail \phi'_i = 
\left(e^{i \Lambda^a T^a}\right)_{ij} \phi_j
\label{phitrsf}
\end{equation}
where $T^a$ are the generators in the chosen representation and $i$
and $j$ are the corresponding indices. In the case of susy QCD these
would be the $\chi$SF containing the quarks, transforming in the
fundamental representation of $SU(3)$, i.e. $i, j \in\{1,2,3\}$. The
Lagrangian then reads 
\begin{equation}
{\cal L} =
\frac{1}{4}\big[ U^a U^a\big]_{\theta\theta} 
+ \frac{1}{4}\big[ \bar{U}^a \bar{U}^a\big]_{\thb\thb}  
+ \big[
\phi_i^\dagger\, \left(e^{2g\, V}\right)_{ij}\, \phi_j 
\big]_{\theta\theta\,\thb\thb}
+ \big[W(\phi_i)\big]_{\theta\theta} 
+ \big[W^\dagger(\phi^\dagger_i)\big]_{\thb\thb} 
\label{susynonabelian}
\end{equation}
where the products of the $\chi$SF $U^a$ and  $\bar{U}^a$ are defined
as in \Eqns{psiprod}{psiBprod}.

The next task is to rewrite \Eqn{susynonabelian} in terms of the
component fields (details are given in Appendix~\ref{app:sample}).
Starting with the first two terms, we note that they take the same
form as \Eqn{FFabelian} with the exception that the normal derivatives
$\partial_\mu$ have to be replaced by the (gauge) covariant
derivatives $D_\mu$, \Eqn{covder}, and the explicit form of
$F^a_{\mu\nu}$ takes the ``non-abelian'' form given in
\Eqn{Fnonabelian}. This can be seen by comparing \Eqn{Unonabelian}
with \Eqn{Uabelian}. Thus the first two terms contain the kinetic
terms of the gluons and gluinos as well as their self interactions due
to the non-abelian nature of the gauge group. Thus susy forces a
non-abelian gluino-gluino-gluon interaction on us through the term
$\sim \lambda \sigma^\mu D_\mu \lambdab$. 

The superpotential terms in \Eqn{susynonabelian} are familiar from the
Wess-Zumino models. This leaves us with the term $\phi^\dagger\,
\left(e^{2g\, V}\right)\, \phi$. Expanding the exponential, the
leading term $\phi^\dagger\, \phi$ is again  familiar from the
Wess-Zumino models and contains the kinetic terms of the
squarks and quarks. The remaining terms, $2 g\, \phi^\dagger\,V\, \phi$
and $2 g^2\, \phi^\dagger\,V\,V\, \phi$ contain the gauge interactions
of the squarks and quarks with the gluons and gluinos.

Putting everything together, the supersymmetric Lagrangian in the
Wess-Zumino gauge for chiral superfields $\phi_i$ (with component
fields $\varphi_i,\ \psi_i$) and vector superfields $V^a$ (with
component fields $v^a_\mu,\ \lambda^a$) for a general gauge group is
given by
\begin{eqnarray}
{\cal L} &=& (D_\mu \varphi)^\dagger_i (D^\mu \varphi)_i 
+ \frac{i}{2} \psi_i \sigma^\mu (D_\mu \psib)_i
- \frac{i}{2} (D_\mu \psi)_i \sigma^\mu \psib_i \nn \\
&-& \frac{1}{4} F^a_{\mu \nu} (F^a)^{\mu \nu}
+  \frac{i}{2} \lambda^a \sigma^\mu (D_\mu \lambdab)^a
- \frac{i}{2} (D_\mu \lambda)^a \sigma^\mu \lambdab^a \nn \\
&-& \sqrt{2} i g \, \psib_i \lambdab^a T^a_{ij} \varphi_j 
+ \sqrt{2} i g \, \varphi^\dagger_i T^a_{ij} \psi_j \lambda^a \nn \\
&-& \frac{1}{2} \frac{\partial^2 W}{\partial \varphi_i \partial
\varphi_j}  \psi_i  \psi_j
- \frac{1}{2} \frac{\partial^2 W^\dagger}{\partial\varphi^\dagger_i
\partial\varphi^\dagger_j} \psib_i \psib_j  
- V(\varphi_i,\varphi^\dagger_j)
\label{masterL}
\end{eqnarray}
The potential is the sum of the $F$-terms, \Eqn{Feliminated}, and
$D$-terms, \Eqn{Deliminated}, and reads
\begin{equation}
V(\varphi_i,\varphi^\dagger_j) = F_i^\dagger F_i + \frac{1}{2} (D^a)^2
= \sum_i \left|\frac{\partial W}{\partial\varphi_i}\right|^2 +
\frac{1}{2} \sum_a (g\, \varphi_i^\dagger\, T^a_{ij}\, \varphi_j +
k^a)^2
\label{potential}
\end{equation}
where the Fayet-Iliopoulos term ${\cal L}_{\rm FI} = 2 \sum_a k^a
[V^a]_{\theta\theta\, \thb\thb}$ can be present only for $U(1)$ gauge
fields.  The most general superpotential $W$ is given by (see
\Eqn{superpotential})
\begin{equation}
W(\varphi_i) = 
a_i\, \varphi_i + \frac{1}{2} m_{ij}\, \varphi_i\varphi_j 
+ \frac{1}{3!} y_{ijk}\,  \varphi_i\varphi_j\varphi_k
\end{equation}
The requirement of gauge invariance imposes constraints on the
coefficients $a_i,  m_{ij}$ and $y_{ijk}$. Finally, the (gauge)
covariant derivatives act as follows:
\begin{eqnarray}
(D_\mu \varphi)_i &=& 
 \partial_\mu \varphi_i + i g\, v_\mu^a T^a_{ij} \varphi_j \nn \\
(D_\mu \psi)_i &=& 
 \partial_\mu \psi_i + i g\, v_\mu^a T^a_{ij} \psi_j  
\label{gaugeCovDer}\\
(D_\mu \lambda)^a &=&  \partial_\mu \lambda^a - g f^{abc} v_\mu^b
\lambda^c \nn
\end{eqnarray}
\Eqn{masterL} is our master equation for the Lagrangian of a susy
gauge theory. Note that at this point we can forget about superfields
and superspace if we want. These concepts have been extremely useful
in deriving \Eqn{masterL}, but are not required any longer once we
have the Lagrangian.

We close this section by looking at the interactions induced by the
various terms of \Eqn{masterL}. Starting with the terms containing
kinetic terms (propagators) we have
\begin{eqnarray}
(D_\mu \varphi)^\dagger_i (D^\mu \varphi)_i &\to&
\begin{picture}(210,30)(0,12)
\SetOffset(10,0)
\DashArrowLine(0,15)(45,15){3}
\SetOffset(75,0)
\Gluon(0,15)(25,15){2.5}{3}
\DashArrowLine(25,15)(45,30){3}
\DashArrowLine(45,0)(25,15){3}
\Vertex(25,15){1.5}
\SetOffset(150,0)
\Gluon(5,30)(25,15){2.5}{3}
\Gluon(5,0)(25,15){2.5}{3}
\DashArrowLine(25,15)(45,30){3}
\DashArrowLine(45,0)(25,15){3}
\Vertex(25,15){1.5}
\end{picture}
\label{phiFR} \\
\frac{i}{2} \psi_i \sigma^\mu (D_\mu \psib)_i +{\rm h.c.}
&\to&
\begin{picture}(210,30)(0,12)
\SetOffset(10,0)
\ArrowLine(0,15)(45,15)
\SetOffset(75,0)
\Gluon(0,15)(25,15){2.5}{3}
\ArrowLine(25,15)(45,30)
\ArrowLine(45,0)(25,15)
\Vertex(25,15){1.5}
\end{picture}
\label{psiFR} \\
- \frac{1}{4} F^a_{\mu \nu} (F^a)^{\mu \nu} &\to&
\begin{picture}(210,30)(0,12)
\SetOffset(10,0)
\Gluon(0,15)(45,15){2.5}{5}
\SetOffset(75,0)
\Gluon(0,15)(25,15){2.5}{3}
\Gluon(25,15)(45,30){2.5}{3}
\Gluon(25,15)(45,0){2.5}{3}
\GCirc(25,16){2.5}{0.8}
\SetOffset(150,0)
\Gluon(5,30)(25,15){2.5}{3}
\Gluon(5,0)(25,15){2.5}{3}
\Gluon(25,15)(45,30){2.5}{3}
\Gluon(25,15)(45,0){2.5}{3}
\GCirc(25,16){2.5}{0.8}
\end{picture}
\label{glueFR} \\
\frac{i}{2} \lambda^a \sigma^\mu (D_\mu \lambdab)^a +{\rm h.c.}
&\to&
\begin{picture}(210,30)(0,12)
\SetOffset(10,0)
\ArrowLine(0,15)(45,15)\Photon(0,15)(20,15){2}{2}\Photon(25,15)(45,15){2}{2}
\SetOffset(75,0)
\Gluon(0,15)(25,15){2.5}{3}
\ArrowLine(25,15)(45,30)\Photon(25,15)(33,21){2}{1}\Photon(37,24)(45,30){2}{1}
\ArrowLine(45,0)(25,15)\Photon(25,15)(33,9){2}{1}\Photon(37,6)(45,0){2}{1}
\GCirc(25,16){2.5}{0.8}
\end{picture}
\label{lambdaFR}
\end{eqnarray}
Dashed lines represent scalars, solid lines superimposed with wavy
lines represent gauginos. The hermitian conjugate of the various
diagrams are not shown. Grey vertices are present only in
non-abelian gauge theories. Turning to the remaining interactions with
no kinetic terms we have
\begin{eqnarray}
- \sqrt{2} i g \, \psib_i \lambdab^a T^a_{ij} \varphi_j +{\rm h.c.}
&\to&
\begin{picture}(210,30)(0,12)
\SetOffset(10,0)
\DashArrowLine(0,15)(25,15){3}
\ArrowLine(25,15)(45,30)
\ArrowLine(25,15)(45,0)\Photon(25,15)(33,9){2}{1}\Photon(37,6)(45,0){2}{1}
\Vertex(25,15){1.5}
\end{picture}
\label{plpFR} \\
\frac{1}{2}  (g\,\varphi_i^\dagger\,T^a_{ij}\,\varphi_j)^2 &\to&
\begin{picture}(210,30)(0,12)
\SetOffset(10,0)
\DashArrowLine(5,30)(25,15){3}
\DashArrowLine(5,0)(25,15){3}
\DashArrowLine(25,15)(45,30){3}
\DashArrowLine(25,15)(45,0){3}
\Vertex(25,15){1.5}
\end{picture}
\label{ppppFR} \\
-\frac{1}{2}\frac{\partial^2 W}{\partial\varphi_i\partial\varphi_j}
 \psi_i  \psi_j+{\rm h.c.}
&\to&
\begin{picture}(210,30)(0,12)
\SetOffset(10,0)
\ArrowLine(0,15)(22.5,15)
\ArrowLine(45,15)(22.5,15)
\Text(22.5,15)[]{$\times$}
\SetOffset(75,0)
\DashArrowLine(0,15)(25,15){3}
\ArrowLine(45,30)(25,15)
\ArrowLine(45,0)(25,15)
\Vertex(25,15){1.5}
\end{picture}
\label{WpsiFR} \\
 \left|\frac{\partial W}{\partial\varphi_i}\right|^2  &\to&
\begin{picture}(210,30)(0,12)
\SetOffset(10,0)
\DashArrowLine(0,15)(22.5,15){2.5}
\DashArrowLine(22.5,15)(45,15){2.5}
\Text(22.5,15)[]{$\times$}
\SetOffset(75,0)
\DashArrowLine(25,15)(0,15){3}
\DashArrowLine(45,30)(25,15){3}
\DashArrowLine(45,0)(25,15){3}
\Vertex(25,15){1.5}
\SetOffset(150,0)
\DashArrowLine(5,30)(25,15){3}
\DashArrowLine(5,0)(25,15){3}
\DashArrowLine(25,15)(45,30){3}
\DashArrowLine(25,15)(45,0){3}
\Vertex(25,15){1.5}
\end{picture}
\label{WphiFR}
\end{eqnarray}

\noindent
The terms introduced through the superpotential are familiar from the
Wess-Zumino model. Indeed, \Eqn{WpsiFR} corresponds to the terms
$-M/2\ \psi\psi$ and $y/2\ \varphi\, \psi\psi$ of \Eqn{WZfinal}
respectively, whereas \Eqn{WphiFR} is responsible for the terms
$-|M|^2 \varphi\varphi^\dagger$,
$-M^*y/2\ \varphi\varphi\varphi^\dagger$ and
$-|y|^2/4\ \varphi\varphi\varphi^\dagger\varphi^\dagger$. The first
terms in \Eqns{WpsiFR}{WphiFR} represent mass terms for the component
fields of the $\chi$SF and the masses have to be equal in a susy
theory.  There are no mass terms for the gauge bosons and the
gauginos. This is to be expected since in an unbroken gauge theory the
gauge bosons are massless. Due to susy, the gauginos have to be
massless as well. To give mass to gauge bosons we have to break gauge
invariance. A simple example is discussed in Section~\ref{sec:Dbreak}.
To give mass to gauginos, we can either break gauge invariance (and
keep susy) such that the gauginos get the same non-vanishing mass as
the gauge bosons, or we can keep gauge invariance (i.e. still have
massless gauge bosons) but break susy. In the MSSM, this is done with
soft breaking terms as will be discussed in
Section~\ref{sec:softbreak}.

\subsection{The unbroken MSSM} \label{sec:MSSM}

With the results of the previous sections we can now go ahead and
write down the susy extension of the Standard Model. We do this by
introducing a $\chi$SF for every fermion of the Standard Model, a VSF
for every gauge boson of the Standard Model and finally two chiral
superfields for the Higgs bosons (the reason for having to introduce
two Higgs superfields will be explained below). By doing this we
introduce the scalar partners of the quarks and leptons, the squarks
and sleptons, and the fermionic partners of the gauge bosons, the
gauginos. We also get a richer Higgs sector, with fermionic
partners. The latter will mix with (some of the) gauginos to produce
the neutralinos and charginos. The $\chi$SF and the VSF are listed in
Tables~\ref{tab:chiral} and \ref{tab:vector} respectively. The
superscripts $\pm$ and $0$ indicate the electric charge $Q_{\rm em}$
with the convention $Q_{\rm em} = T_3 + Y$, where $T_3$ is the third
component of isospin.

\medskip
\begin{table}[h]
\begin{center}
\begin{tabular}{|l|c|c|c|c|}
\hline\rule{0cm}{0.5cm}
\phantom{$\biggl( \biggr)$}
 & LH$\chi$SF & spin 0 & spin $\frac{1}{2}$ & $(SU(3), SU(2), U_Y(1))$ \\
\hline\hline\rule{0cm}{0.5cm}
squarks and quarks & $Q$ & $(\tilde{u}_L,\tilde{d}_L)$ & $(u_L,d_L)$ &
$(3,2,\frac{1}{6})$ 
 \\ \rule{0cm}{0.5cm}
 & $U$ & $\tilde{u}_R^\dagger$ & $u_R^\dagger$ & $(\bar{3},1,-\frac{2}{3})$
 \\ \rule{0cm}{0.5cm}
\phantom{$\biggl( \biggr)$} 
& $D$ & $\tilde{d}_R^\dagger$ & $d_R^\dagger$ & $(\bar{3},1,\frac{1}{3})$ \\
\hline\rule{0cm}{0.5cm}
sleptons and leptons & $L$ & $(\tilde{\nu},\tilde{e}_L)$ & $(\nu,e_L)$
 & $(1,2,-\frac{1}{2})$ 
\\ \rule{0cm}{0.5cm}  
\phantom{$\biggl( \biggr)$} 
& $E$ & $ \tilde{e}_R^\dagger$ & $e_R^\dagger $ & $(1,1,1)$ \\
\hline\rule{0cm}{0.5cm}
higgs and higgsinos & $H_u$ & $(h_u^+, h_u^0)$ & 
 $(\tilde{h}_u^+,\tilde{h}_u^0)$ &  $(1,2,\frac{1}{2})$ 
\\ \rule{0cm}{0.5cm}
\phantom{$\biggl( \biggr)$} 
& $H_d$ & $(h_d^0, h_d^-)$ & 
 $(\tilde{h}_d^0, \tilde{h}_d^-)$ &  $(1,2,-\frac{1}{2})$ 
\\
\hline\hline
\end{tabular}
\end{center}
\ccaption{}{Chiral superfields of the MSSM with their particle
  content. The transformation property under $SU(3)\times SU(2)$ and
  the value of $U_Y(1)$ is given in the last column. There are three
  copies of the quark and lepton superfields, one for each family.
\label{tab:chiral} }
\end{table}

\begin{table}[h]
\begin{center}
\begin{tabular}{|l|c|c|c|c|}
\hline\rule{0cm}{0.5cm}
\phantom{$\biggl( \biggr)$}
 & VSF & spin $\frac{1}{2}$ & spin 1 & $(SU(3), SU(2), U_Y(1))$ \\
\hline\hline\rule{0cm}{0.5cm}
gluinos and gluons & $G$ &  $\tilde{g}$ & $g$ & $(8,1,0)$ \\
\hline \rule{0cm}{0.5cm}
winos and $W$-bosons & $W$ & $ \widetilde{W}^\pm, \widetilde{W}^0 $ &
$W^\pm, W^0$ &  
 $(1,  3, 0)$ \\ \hline \rule{0cm}{0.5cm}
bino and $B$-boson & $B$ & $\widetilde{B} $ & $ B$ & 
 $(1,  1, 0)$ \\ \hline\hline
\end{tabular}
\end{center}
\ccaption{}{Vector superfields of the MSSM with their particle
  content. The transformation property under $SU(3)\times SU(2)$ and
  the value of $U_Y(1)$ is given in the last column.
\label{tab:vector} }
\end{table}

It is clear that constructing such a theory by using \Eqn{masterL}
will result e.g. in squarks and sleptons with the same mass as the
corresponding quarks and leptons. Since this is in clear contradiction
to observation, we will have to find a way to break susy to make the
model phenomenologically acceptable. This issue will be addressed in
Section~\ref{sec:Breaking}. Here we focus on the simpler task of
writing down the strictly susy extension of the Standard Model.

Following \Eqn{masterL} we see that after having fixed the list of
$\chi$SF and VSF, i.e. the matter fields and the gauge group, the only
freedom we have is in choosing the superpotential $W(\phi_i)$. This
completely fixes the Lagrangian. As stated repeatedly, we have to
make sure that $W$ is gauge invariant and that it is an analytic
function of the LH$\chi$SF. It is for this reason that in
Table~\ref{tab:chiral} we have listed all $\chi$SF as LH$\chi$SF,
i.e. we take the hermitian conjugate of the right-handed fields to
obtain a LH$\chi$SF.

Let us start with a term in the superpotential, $W_1$, that gives rise
to down-type quark masses. As in the Standard Model this is done by
coupling the quark fields to a Higgs field with a non-vanishing vacuum
expectation value (vev). The term is given by
\begin{equation}
W_{1}(\phi_i) = -\, y_d\, D\, Q\, H_d \equiv 
-\, \left(y_d\right)_{f_i f_j} D^{f_i}\, 
Q^{f_j}_a \epsilon^{a b} \left(H_d\right)_b
\label{Dmassterm}
\end{equation}
and is usually written as in the l.h.s. of \Eqn{Dmassterm}. On the
r.h.s. we have introduced (nearly) all labels. First, $f_i,\,
f_j\in\{1,2,3\}$ label the family/flavour. Second, $a,\, b
\in\{1,2\}$ are $SU(2)$ labels. The $\epsilon^{a b}$ is needed to make
the term $Q\, H_d$ a singlet under $SU(2)$. Since $D$ is also a
singlet under $SU(2)$ the whole term is gauge invariant with respect
to $SU(2)$. The gauge invariance with respect to $SU(3)$ is trivial
(which is why we omitted colour labels on the r.h.s. of
\Eqn{Dmassterm}), since $ 3 \times \bar 3 = 1 + 8$ contains a singlet
and $H_d$ is a singlet. The hypercharges of the three $\chi$SF add to
zero, thus the term is indeed gauge invariant under $SU(3)\times
SU(2)\times U_Y(1)$.

Giving the Higgs a non vanishing vev then results in a mass term
for the down-type quarks. More precisely, writing the term
\Eqn{Dmassterm} in terms of its scalar component fields, as required
for \Eqn{masterL}, we get 
\begin{equation}
W_{1}(\varphi_i) = 
-\, \left(y_d\right)_{f_i f_j}\, (\tilde{d}_R^\dagger)^{f_i} 
\left((\tilde{u}_L)^{f_j} \, h_d^- - 
  (\tilde{d}_L)^{f_j} \, h_d^0 \right)
\end{equation}
If the neutral component of the Higgs gets a vev, $\langle h_d^0
\rangle = v_d$, we obtain a mass term for the fermions through
the term
\begin{equation}
- \frac{1}{2} \frac{\partial^2 W_1}{\partial \varphi_i \partial
\varphi_j}\,  \psi_i  \psi_j + {\rm h.c.} \Rightarrow
- \frac{1}{2}\, v_d \left(y_d\right)_{f_i f_j}\, 
(d_R^\dagger)^{f_i}\, d_L^{f_j} + {\rm h.c.}
\end{equation}
where on the l.h.s. we have given the general expression as in
\Eqn{masterL} and on the r.h.s. the explicit expression we obtain from
$W_1$ as given in \Eqn{Dmassterm} with $\varphi_i = (\tilde
d_R^\dagger)^{f_i}$, $\varphi_j = (\tilde d_L)^{f_j}$, $\psi_i =
(d_R^\dagger)^{f_i}$ and $\psi_j = d_L^{f_j}$. Thus we have a mass
matrix in family space, $m_{f_i f_j} = v_d (y_d)_{f_i f_j}$ which we
have to diagonalize to obtain the masses of the three down-type
quarks. The squarks obtain their mass from the term
\begin{equation}
-\sum_i\left|\frac{\partial W_1}{\partial \varphi_i}\right|^2
\Rightarrow \
-\, v_d^2\, |(y_d)_{f_i f_j}|^2\, \left(
\tilde{d}^{f_i}_L (\tilde{d}_L^{f_j})^\dagger + 
   \tilde{d}^{f_i}_R (\tilde{d}^{f_j}_R)^\dagger \right)
\end{equation}
which results in the same masses for the squarks and quarks. Note that
both, the squarks and quarks get their masses from a non-zero vev of
the scalar component field of the neutral Higgs boson. Charged fields
or fermionic fields cannot get a vev without violating charge
conservation or Lorentz invariance.

Of course, there are more terms associated with the superpotential
term $W_1$. If we insert $W_{1} = -\, y_d\, D\, Q\, H_d$ into
\Eqn{masterL} we get interactions of Higgs bosons with fermions e.g
$h^0\to d\, \bar{d}$ and $h^- \to \bar{u}\, d$ or interactions of
squarks with higgsinos and quarks, e.g. $\tilde{d} \to \tilde{h}^-\,
u$. These interactions correspond to those exemplified in
\Eqn{WpsiFR}. There are also four-point scalar interactions such as
$\tilde{d}\tilde{u}\to \tilde{d}\tilde{u}$ as shown in
\Eqn{WphiFR}. The higgsinos actually mix with the fermionic partners
of gauge bosons to form charginos and neutralinos.  For a more
complete discussion and a list of interactions with Feynman rules we
refer to Refs.~\cite{Haber:1984rc,Rosiek:1989rs}.

The charged leptons obtain their mass in exactly the same way, i.e. by
introducing the term $W_2 = -\, y_e\, E\, L\, H_d$. The $SU(2)$
doublets are combined as in \Eqn{Dmassterm} to obtain a gauge
invariant term. The gauge invariance with respect to $SU(3)$ and
$U_Y(1)$ is obvious. Giving mass to the up-type quarks is not so
easy. In the Standard Model, this is done with the same Higgs boson,
by introducing a term $\sim U Q H^\dagger$. However, this term
violates susy, because it contains $H^\dagger$ and therefore the
superpotential is not an analytic function of the LH$\chi$SF any
longer. Thus we have no other choice than to introduce a second Higgs
doublet $H_u$ with the neutral component field that gets a vev in the
$T_3=-1/2$ position of the doublet, $\langle h_u^0\rangle = v_u$. Then
we can write the gauge invariant term $W_3 = y_u\, U\, Q\, H_u$ which
gives a mass to the up-type quarks. The presence of the second Higgs
doublet also ensures the cancellation of anomalies.

Having a second Higgs doublet allows us to construct another gauge
invariant term, $W_4 = \mu\, H_u\, H_d$, such that the MSSM
superpotential reads
\begin{equation}
W_{\rm MSSM}=  y_u\, U\, Q\, H_u
- y_d\, D\, Q\, H_d -  y_e\, E\, L\, H_d + \mu \, H_u\, H_d
\label{Wmssm}
\end{equation}
These are all the terms we want but, most unfortunately, not all the
terms we get. There are many more gauge invariant terms that can be
included in the superpotential and, unless there is a good reason to
leave them out, from a theoretical point of view we have to include
them.

Looking at Table~\ref{tab:chiral} we see that the following terms are
also all gauge invariant
\begin{equation}
W_{\not R} = \frac{1}{2} \lambda\, E\, L\, L 
+ \lambda'\, D\, L\, Q + \mu'\, L\, H_u
+\frac{1}{2} \lambda''\, U\, D\, D
\label{Rviol}
\end{equation}
The factors 1/2 are introduced to account for the symmetry. The gauge
invariance under $SU(3)$ of the term $U\, D\, D$ implies that we have
to take the completely antisymmetric colour combination,
i.e. $\epsilon_{ijk}\, U^i\, D^j\, D^k$, where $i$, $j$ and $k$ are
colour indices. Thus this is the same colour combination as e.g. in a
antiproton. Note that the gauge invariant term $E\, H_d\, H_d$
vanishes due to the $\epsilon^{ab}$ in the combination of the two weak
doublets, as detailed in \Eqn{Dmassterm}. We also remind the reader
that terms with more than three $\chi$SF lead to a non-renormalizable
theory and therefore are left out. The problem with the terms in
\Eqn{Rviol} is that they violate lepton number (the first three terms)
and baryon number (the last term). This leads to serious problems with
proton decay (see e.g. Ref~\cite{Martin:1997ns}).

These problems can be avoided by pulling another symmetry out of a
hat. Usually this is R-parity, a multiplicative quantum number defined
in terms of baryon number $B$, lepton number $L$ and spin $s$ as
$R\equiv (-1)^{3B+L+2s}$ such that Standard Model particles (including
the Higgs bosons) have $R=1$, whereas all superpartners have
$R=-1$. Note that the various component fields of a superfield have
different R-parity due to the spin contribution. Thus we cannot
associate R-parity to a superfield and it is not immediately obvious
that the terms in \Eqn{Rviol} violate R-parity. From this point of
view a more convenient symmetry is matter parity, defined as
$(-1)^{3B+L}$. Due to angular momentum conservation matter parity
conservation and R-parity conservation are equivalent. The former has
the advantage that it is defined for a superfield. The lepton and
quark superfields have matter parity $-1$, whereas the Higgs and
vector superfields have matter parity $+1$. Keeping in mind that this
is a multiplicative quantum number, it is now immediately obvious that
all terms in \Eqn{Wmssm} have matter parity $+1$, whereas all terms in
\Eqn{Rviol} have matter parity $-1$.

Another option to avoid problems with proton decay is to impose baryon
or lepton number conservation, leading to R-parity violating
scenarios. In either case, it is disturbing that in the MSSM an
additional symmetry has to be introduced to avoid these problems. In
the Standard Model, such problematic terms are absent accidentally,
i.e. without any further requirements.

\section{Breaking supersymmetry } \label{sec:Breaking}

The MSSM Lagrangian of Section~\ref{sec:MSSM} leads to superpartners
with the same mass as the corresponding Standard Model particles.
Obviously this is not in accord with Nature and therefore not
acceptable. Thus, we have to break susy in such a way as to give the
superpartners a larger mass. It is also clear that we must not break
susy by brute force. The situation is similar to the case of gauge
theories, where gauge symmetry implies massless gauge bosons in
contrast with experiment. As is well known, this problem can be solved
by breaking gauge symmetry spontaneously, i.e. the Lagrangian is still
gauge invariant but the ground state of the theory does not share this
symmetry. This gives mass to the $W$ and $Z$ bosons while maintaining
the wanted features of the symmetry. We want to do the same for susy.

Before we look at the various possibilities explicitly, let us make a
few general considerations. Let us start with \Eqn{algQQ} and multiply
it by $(\sigmab^0)^{\bd\alpha}$. On the l.h.s. we use the fact that
$(\sigmab^0)^{\bd\alpha}$, as defined in \Eqn{app:pauli}, is simply
the unit $2\times 2$ matrix. On the r.h.s. we use \Eqn{app:tr} and
thus obtain
\begin{equation}
Q_1 Q_1^\dagger +Q_1^\dagger Q_1 + Q_2 Q_2^\dagger +Q_2^\dagger Q_2 = 
4\, g^{0\mu}P_\mu = 4\, P^0 = 4\, H
\label{Halg}
\end{equation}
where $H$ is the Hamiltonian and we used $(Q_\alpha)^\dagger =
\bar{Q}_\ad$. From \Eqn{Halg} we see that susy theories have the
remarkable property that $H$ is bounded from below, i.e. for any state
$|b\rangle$ we have $\langle b|H|b\rangle \ge 0$. 

Let us specialize to the ground state $|0\rangle$ of our theory. If
susy is not spontaneously broken, the ground state shares the symmetry
of the Lagrangian, i.e. $|0\rangle$ is invariant under susy. This
means $S(0,\zeta,\zeb)|0\rangle = |0\rangle$, with $S(0,\zeta,\zeb)$
as given in \Eqn{susytrsf}, which entails $Q_\alpha|0\rangle = 0$ and
$\bar Q_\ad|0\rangle = (Q_\alpha)^\dagger|0\rangle = 0$. From
\Eqn{Halg} we then immediately conclude $\langle 0|H|0\rangle =
0$. Again this is a remarkable property of unbroken susy. Compare this
for example to the normal harmonic oscillator, where the ground state
energy is $1/2\neq 0$. In a susy harmonic oscillator, the fermionic
part cancels this contribution and the ground state energy is zero.

If susy is spontaneously broken, the ground state does not share the
symmetry of the Lagrangian, i.e. $|0\rangle$ is not invariant under
susy. This implies that $Q_\alpha |0\rangle \neq 0$ and thus $\langle
0|H|0\rangle > 0$. This is the crucial criteria for the construction
of spontaneously broken susy. 

In order to obtain a strictly positive ground state energy the
potential $V$ has to satisfy $V|_{\rm min}>0$. According to
\Eqn{potential} the potential has two terms, a F-term given by $V_F =
\left|{\partial W} / {\partial\varphi_i}\right|^2$ and a D-term, $ V_D
= 1/2\, (g\,\varphi_i^\dagger\, T^a_{ij}\, \varphi_j + k^a)^2$ and
obviously satisfies $V\ge 0$ in agreement with $\langle 0|H|0\rangle
\ge 0$. If we want spontaneous symmetry breaking we either need
$V_F|_{\rm min} >0$ ({\it F-term breaking}) or $V_D|_{\rm min} >0$
({\it D-term breaking}) or a combination of both. We will look at
explicit examples of the two cases in turn.

\subsection{F-term breaking} \label{sec:Fbreak}

The canonical example of F-term breaking is the O'Raifeartaigh (OR)
model~\cite{O'Raifeartaigh:1975pr}.  Consider the case where we have
three $\chi$SF and the superpotential
\begin{equation}
W_{\rm OR}(\phi_i)=-a\, \phi_1 + m\, \phi_2 \phi_3 + \frac{y}{2} \phi_1
\phi_3^2
\label{Wor}
\end{equation}
The potential is then given by
\begin{equation}
V_{\rm OR}  =  \sum_i
 \left|\frac{\partial W_{\rm OR}(\varphi_i)}{\partial\varphi_i}\right|^2
= \left|a-\frac{y}{2} \varphi_3^2\right|^2 + 
\left|m\, \varphi_3 \right|^2+ 
\left|m\, \varphi_2 +y\,\varphi_1 \varphi_3  \right|^2
\label{Vor}
\end{equation}
Looking at the first two terms of $V_{\rm OR}$ we conclude $V_{\rm
  OR}>0$, which is precisely what we want. The potential has three
extrema. If we assume $a<m^2/y$, the absolute minimum of the potential
is at $\varphi_2=\varphi_3 = 0$ and arbitrary $\varphi_1$. In this
case $V_{\rm OR}|_{\rm min} = a^2\neq 0$.

To verify that susy has been broken, let us compute the masses of the
fermions and scalars in this theory. For each of the three $\chi$SF we
have two real scalars $\varphi_i^{\rm Re}$ and $\varphi_i^{\rm Im}$
and a Weyl spinor $\psi_i$, i.e. two real bosonic and two real
fermionic degrees of freedom. To compute the fermion mass, we first
obtain the mass matrix
\begin{equation}
- \frac{1}{2} 
\frac{\partial^2 W_{\rm OR}}{\partial \varphi_i \partial\varphi_j}\,  
\psi_i  \psi_j  + {\rm h.c.}=
- \frac{1}{2} \left( 2 m\, \psi_2 \psi_3 + 
 y \langle\varphi_1\rangle\, \psi_3\psi_3 \right)+ {\rm h.c.}
\label{ORfermionM}
\end{equation}
These are all bilinear in $\psi_i$ terms we get for
$\langle\varphi_2\rangle =\langle\varphi_3 \rangle = 0$ and
$\langle\varphi_1\rangle\neq 0$. There is no mass term at all for
$\psi_1$, resulting in a massless fermion. This is not surprising. We
know from gauge theories that spontaneous breaking of a global
(bosonic) symmetry results in a massless Goldstone boson. Here we have
the spontaneous breaking of global susy, a fermionic symmetry, thus we
get a massless Goldstone fermion, usually called {\it goldstino}.
Linear combinations of the other two fermions, $\psi_2$ and $\psi_3$
have mass $m$.

Let us now compute the mass of the scalars. To do this we expand the
potential around $\varphi_i = 0$ and consider the bilinear terms in
$\varphi_i^{\rm Re}$ and $\varphi_i^{\rm Im}$. We get two massless
scalars, $\varphi_1^{\rm Re}$ and $\varphi_1^{\rm Im}$ and two
scalars of mass $m$, $\varphi_2^{\rm Re}$ and $\varphi_2^{\rm
  Im}$. This is still completely susy, as these masses agree with the
corresponding fermion masses. However, the breaking of susy manifests
itself in the remaining scalar masses, $\varphi_3^{\rm Re}$ and
$\varphi_3^{\rm Im}$ which are found to be $\sqrt{m^2-a\, y}$ and
$\sqrt{m^2+a\, y}$. Thus the scalar masses differ from the masses of
the corresponding fermions, a clear sign that susy is broken.

The problem with this mechanism is that it does not provide what we
want from a phenomenological point of view. We would like to break
susy such that all of the (yet undiscovered) scalars get a larger
mass than the fermions. In the example above, one of the scalars has a
higher mass than the corresponding fermion, the other has a lower
mass. Unfortunately, this is a general feature~\cite {Ferrara:1979wa}
and can be written as
\begin{eqnarray}
{\rm STr}\, M^2 \equiv \sum (-1)^s (2s+1)\, m_s^2 = 0
\label{str}
\end{eqnarray}
In the above relation the {\it supertrace} sums over all component
fields, $s$ denotes the spin and $m_s$ is the mass associated with the
real component field of spin $s$. This implies that with this
mechanism we will always get a symmetric shift in the masses,
i.e. some superpartners are heavier and others have smaller mass than
the Standard Model particles such that the average mass remains the
same.

It is important to note that this relation holds only at tree level
and is in general violated by loop corrections. This does not help
directly, as loop corrections will never be able to shift e.g. the
selectron mass from below the electron mass to something like
100~GeV. But it does leave a window for F-term spontaneous susy
breaking. If we have F-term breaking not directly in the MSSM, but in
a hidden sector, then it is possible to mediate the susy breaking by
loop effects into the MSSM and avoid the constraint of \Eqn{str}.

\subsection{D-term breaking} \label{sec:Dbreak}

Let us come back to the abelian susy gauge theory discussed in
Section~\ref{sec:QED}. For simplicity we assume there is only one
$\chi$SF and the superpotential vanishes, $W(\phi)=0$. However, we
have a Fayet-Iliopoulos term. After eliminating the $D$ component
field the Lagrangian reads
\begin{eqnarray}
{\cal L} &=&
\frac{1}{4}\big[ U^\alpha U_\alpha\big]_{\theta\theta} 
+ \frac{1}{4}\big[ \bar U_\ad \bar U^\ad\big]_{\thb\thb} 
+  \big[
\phi^\dagger\, e^{2 g\, V}\, \phi \big]_{\theta\theta\, \thb\thb}
+\big[2 k\, V\big]_{\theta\theta\, \thb\thb} \nn \\
&=& - \frac{1}{4} F^{\mu\nu}F_{\mu\nu} 
+ (D_\mu \varphi)^\dagger (D^\mu \varphi) 
-\frac{1}{2} \left(g\, \varphi^\dagger \varphi+k\right)^2 \nn \\
&& -\ \left(\frac{i}{2} (\partial_\mu\lambda)\sigma^\mu\lambdab +
  \frac{i}{2} (D_\mu \psi) \sigma^\mu \psib +
  \sqrt{2} i g \, \psib \lambdab\,  \varphi + {\rm h.c.}\right)
\label{LDbreak}
\end{eqnarray}
with $D^\mu = \partial^\mu+ig\, v^\mu$.

Let us focus on the potential $V=D^2/2=(g\, \varphi^\dagger
\varphi+k)^2/2$ which holds the key to spontaneous breaking of
susy. We would like $V$ to be strictly positive $V|_{\rm min}>0$. In
order to see whether we can achieve this we have to distinguish two
cases. Either the scalar field $\varphi$ gets a vev or it does not.

Starting with the first scenario we see that the presence of $k$ does
not prevent $D=0$. What can happen is that $\varphi$ gets a vev such
that $V|_{\rm min}= 0$, i.e. $\langle\varphi^\dagger\varphi\rangle =
-k/g>0$. Thus what we actually achieve is not spontaneous breaking of
susy but rather spontaneous breaking of gauge invariance. Indeed, the
term $(D_\mu \varphi)^\dagger (D^\mu \varphi)$ will result in a gauge
boson mass term $g^2 v^\mu v_\mu \langle \varphi^\dagger \varphi
\rangle = -k\, g\, v^\mu v_\mu = (m_v^2/2)\ v^\mu v_\mu$, i.e the
gauge boson mass is $m_v=\sqrt{-2 k\, g}$.  The additional degree of
freedom associated with the mass of the gauge boson comes from one of
the scalars. This can be seen by writing $\varphi = (\varphi^{\rm Re}
+ i\, \varphi^{\rm Im})/\sqrt{2}$ and expanding the potential around
$\langle\varphi^{\rm Im}\rangle = \sqrt{ -2\, k/ g}$. The scalar field
$\varphi^{\rm Im}$ gets a mass term $k\,g\,(\varphi^{\rm Im})^2 =
-(m_\varphi^2/2)\, (\varphi^{\rm Im})^2$, i.e. the same mass as the
gauge boson. However, the other scalar field, $\varphi^{\rm Re}$ does
not get a mass term. This is the Goldstone boson that gets absorbed by
the initially massless gauge boson. The fermion $\psi$ of the $\chi$SF
and the gaugino $\lambda$ form a Dirac spinor $\Psi =
(\psi,\lambdab)$, as in \Eqn{DiracToWeyl}, and also get a mass term
from $-\sqrt{2}i\, g\, \psib \lambdab\, \langle\varphi \rangle + {\rm
  h.c} = \sqrt{-2g\,k}\, \overline{\Psi}\Psi$. In fact the mass of the
gauge boson is the same as the gaugino mass as it has to be since susy
is not broken. For the same reason, the (massive) scalar and the other
fermion also have the same mass. This scenario is simply the susy
generalization of the Higgs mechanism, where a massless $\chi$SF and a
massless VSF combine to a massive VSF.

To achieve what we set out for we have to prevent $D=0$. If
$\langle\varphi\rangle = 0$ we have $V|_{\rm min}=k^2/2 \neq 0$. The
gauge boson $v_\mu$, its partner the gaugino $\lambda$ as well as the
fermion $\psi$ of the $\chi$SF all remain massless. The only particle
that gets a mass is the scalar, through the term $ -g\, k\,
\varphi^\dagger\varphi$ from the potential. This corresponds to a mass
$m_\varphi=\sqrt{g\,k}$ for the two (real) scalar fields $\varphi^{\rm
  Re}$ and $\varphi^{\rm Im}$.

That is precisely what we wanted to achieve! Thus the key for D-term
susy breaking is to prevent the scalar fields to develop a vev.  We
can achieve this by giving the scalar fields large masses through
superpotential terms. Therefore we now consider a non-vanishing
superpotential. To get a gauge invariant superpotential we need a pair
of $\chi$SF, $\phi_i,\ i\in \{1,2\}$ with opposite charges $q_i$ with
respect to the $U(1)$ under consideration. More precisely, the fields
have to have gauge transformations like $\phi_i \rightarrowtail e^{-i
  q_i\Lambda}\phi_i$ with $q_1=-q_2$. This enables us to write a gauge
invariant term $W=m\, \phi_1\phi_2$ in the superpotential.  The scalar
potential then also gets a F-term contribution and reads
\begin{equation}
V = |m|^2 \sum_{i=1}^2 \varphi_i^\dagger\varphi_i + 
\frac{1}{2} \Big(k+ g \sum_{i=1}^2
q_i\, \varphi_i^\dagger\varphi_i\Big)^2 
\label{VFD}
\end{equation}
with $\varphi_i^\dagger\varphi_i = 1/2\, ((\varphi_i^{\rm Re})^2 +
(\varphi_i^{\rm Im})^2)$. If we choose $|m|^2$ large enough, $|m|^2 >
g |q_i| k$, the minimum of the potential is at $\langle \varphi_i^{\rm
  Re} \rangle = \langle \varphi_i^{\rm Im} \rangle = 0$ and we have
$V|_{\rm min}=k^2/2 \neq 0$. 

Let us try to apply this mechanism to the MSSM with $U_Y(1)$ as the
abelian group. Now we immediately face a problem. We can give large
masses to the Higgs scalars through the superpotential term $\mu\,
H_u\, H_d$ but not to the other scalars. There are no gauge invariant
terms corresponding to $W=m\, \phi_1\phi_2$ in the MSSM superpotential
given in \Eqns{Wmssm}{Rviol}. Thus, however nice the D-term susy
breaking mechanism is, it cannot be applied to the MSSM. What would
happen is that e.g. the squark fields develop a vev, rather than susy
being broken. This is not acceptable as it would break electric charge
and colour conservation, the last thing we want.

As for F-term breaking, in order for D-term breaking to be
phenomenologically acceptable, it would have to happen in a hidden
sector, with a new $U(1)$ group. The breaking then would have to be
mediated to the visible sector, the MSSM.

Let us close this section with a remark concerning the supertrace
formula \Eqn{str}. In our initial D-term breaking example with only one
$\chi$SF the two real scalars of the $\chi$SF obtain a mass shift,
whereas all other particles remain massless. This clearly violates the
supertrace formula. In fact, \Eqn{str} can be generalized by writing
the r.h.s. as $\sum q_i^2 g\, \langle D\rangle$. However, as we have
seen, for a realistic (gauge invariant) example we need the $\chi$SF
to come in pairs with opposite charges. Thus for every mass shift
$\delta m_{\varphi_1} = q_1\, \sqrt{g\,k}$ of a scalar component field
we get an opposite mass shift $\delta m_{\varphi_2}=q_2\, \sqrt{g\,k}
= - \delta m_{\varphi_1}$ and \Eqn{str} holds again.

\subsection{Soft breaking and the hierarchy problem} \label{sec:softbreak}

In the previous two sections we have seen that while it is possible to
break susy spontaneously either through F-term or D-term breaking,
neither option works directly for the MSSM. The standard procedure
then is to introduce a hidden sector, break susy in the hidden sector
and mediate the breaking to the visible sector, the MSSM, either
through gravity, gauge interactions or by other means. If we did know
the details of the hidden sector and the mediation we could compute
the induced breaking in the visible sector. Sadly, we don't. Thus we
have to parameterize our ignorance. If we choose the latter option we
introduce susy breaking terms by hand. The idea is to measure these
parameters and hopefully, once a consistent picture arises, to infer
from these measurements the theory behind susy breaking.

Inserting susy breaking terms by hand we have to be careful not to
destroy all the nice features of susy. One of these features is the
much celebrated cancellation of quadratic divergences and its relation
to the hierarchy problem. 

To understand this let us start by considering a fermion, say the
electron, and recapitulate some basic properties about
renormalization. In the Lagrangian we have a term $m_0\,
\overline{\Psi} \Psi$, where $m_0$ is the bare mass. The parameter
$m_0$ is related in a particular way (depending on the precise
definition of the mass) to the (renormalized) theoretical mass $m_{\rm
  th}$. At tree level, we have $m_0 = m_{\rm th}$, at one loop we have
$m_{\rm th} = m_0 + \delta m$, where the one-loop corrections $\delta
m \sim \alpha\, m_0\, K$. Here $\alpha$ is the (electromagnetic)
coupling and $K$ a calculable coefficient, depending on the
regularization and precise definition of the mass. Due to the presence
of ultraviolet singularities in loop integrals, $\delta m$ is actually
divergent. If we use dimensional regularization in $D=4-2\epsilon$
dimensions, $\delta m$ contains a pole $1/\epsilon$.  The physical
reason for this divergence is the breakdown of our field-theory
picture at large energies because for instance it does not include
gravity. In order to be able to proceed in our field theory approach
we absorb our ignorance into a counterterm and relate it to an
experimentally measured value, in our case, the electron mass $m_{\rm
  exp}$. Thus we set $m_{\rm th} = m_0 + \delta m = m_{\rm exp}$ and
thereby determine $m_0$. Once we have done this for the electron mass
(and a few more quantities) we can then go and predict any other
quantity within our field theory approach.

Since $m_{\rm th}$ is finite but $\delta m$ is divergent, we know
$m_0$ has to be divergent as well.  Thus we have an infinite fine
tuning in that two infinite quantities, $m_0$ and $\delta m$, conspire
to give a value $m_{\rm th} = m_0 + \delta m = 0.5$~MeV (in the case
of the electron). Due to the above mentioned reason i.e. our accepted
ignorance of what is happening at very large energy scales, nobody is
worried about this. However, we certainly want our theory to be valid
up to a certain scale $\Lambda$. Thus, if we replace the usual
dimensional regularization by a more physical regularization, which
consists of introducing a cutoff $\Lambda$ in our loop integrals, we
would hope not to have this fine tuning problem.

In the case of the electron, or any fermion, this is the
case. Considering the power counting of the one-loop diagram that
contributes to $\delta m_{F}$, the correction to the fermion mass
$m_{F}$, we get from the fermion and photon propagator four powers of
the integration momentum $k$ in the denominator and one in the
numerator.
\begin{equation}
\begin{picture}(80,25)(0,10)
\Line(0,5)(70,5)
\PhotonArc(35,5)(20,0,180){2}{8.5}
\Vertex(15,5){1.5}\Vertex(55,5){1.5}
\end{picture}
\Rightarrow\ 
\delta m_{F} \simeq 
\alpha\, \int^{\Lambda} d^4 k\, \frac{\{k\}}{k^2\, (k^2-m_F^2)} 
\simeq \alpha\, m_{F} \log\frac{\Lambda}{m_F}
\label{fermionSE}
\end{equation}
This seems to lead to a linear divergence in $\Lambda$, i.e. $\delta
m_{F} \sim \alpha \Lambda$. However, the linear term in the numerator
always vanishes upon integration, as indicated by the curly brackets,
if our regulator does not break Poincar\'e invariance. Thus we are
left with only a logarithmic divergence $\delta m_F \sim \alpha\, m_F
\log(\Lambda/m_{F})$. As a result even for very large values of
$\Lambda\sim~10^{15}$~GeV, we have $\delta m_{F} \lesssim m_{F}$,
i.e. the correction is of the order of $m_{F}$ and there is no fine
tuning required.

Let us now repeat this exercise for any gauge boson. In general, we
have two kinds of one-loop diagrams contributing to $\delta m_G^2$,
fermion loops and gauge boson loops and we generically denote the
masses of particles in the loop by $m_L$. In both cases, we get four
powers of $k$ in the denominator and two in the numerator. Those in
the numerator are either from the fermion propagator or the
gauge-boson interaction vertices.
\begin{equation}
\begin{picture}(160,25)(0,4)
\Gluon(0,5)(20,5){3}{2}\Gluon(50,5)(70,5){3}{2}
\GlueArc(35,5)(15,0,360){2}{12}
\Vertex(20,5){1.5}\Vertex(50,5){1.5}
\Text(78,7)[]{$+$}
\SetOffset(85,0)
\Gluon(0,5)(20,5){3}{2}\Gluon(50,5)(70,5){3}{2}
\ArrowArc(35,5)(15,0,180)
\ArrowArc(35,5)(15,180,360)
\Vertex(20,5){1.5}\Vertex(50,5){1.5}
\end{picture}
\Rightarrow\ 
\delta m_G^2 \simeq 
\alpha \int^{\Lambda}\!\! d^4 k\, \frac{\{k^2\}}{k^2\, k^2} 
\simeq \alpha\, m_{L}^2 \log\frac{\Lambda}{m_{L}}
\label{gaugeSE}
\end{equation}
From power counting we would expect a quadratic divergence in $\delta
m_G^2$. However, as indicated by the curly brackets, there is a
cancellation of these quadratic singularities and the final answer is
only logarithmically divergent. As for $\delta m_F$, even for very
large values of $\Lambda$ we have $\delta m_G^2$ and $m_G^2$ of the
same order and no fine tuning is required. Thus our theory potentially
could be valid up to very large energy scales.

The cancellation of quadratic singularities is not a coincidence. It
is a symmetry that ensures this cancellation. In an unbroken gauge
theory the gauge boson remains massless to all orders, so the
cancellation is actually even stronger. Not only do the quadratic
singularities cancel, but $\delta m^2_G=0$. In a spontaneously broken
gauge theory this is no longer the case, but gauge symmetry still
ensures the cancellation of the quadratic singularities. Also in the
case of the fermion there is a symmetry that protects the fermion mass
from large corrections. In fact, if the fermion is initially massless
there is an additional symmetry, chiral symmetry, which prevents the
generation of a non-vanishing mass.

So far so good, but what about scalars, i.e. the Higgs boson in the
Standard Model. Let us consider the correction $\delta m^2_S$ due to a
fermion loop. Again, we expect two powers of $k$ in the numerator and
four powers of $k$ in the denominator.
\begin{equation}
\begin{picture}(80,25)(0,4)
\DashLine(0,5)(20,5){2}\DashLine(50,5)(70,5){3}
\ArrowArc(35,5)(15,0,180)
\ArrowArc(35,5)(15,180,360)
\Vertex(20,5){1.5}\Vertex(50,5){1.5}
\end{picture}
\Rightarrow\ 
\delta m_S^2 \simeq 
\alpha \int^{\Lambda}\!\! d^4 k\, \frac{k^2}{(k^2-m_L^2)\, (k^2-m_L^2)} 
\simeq \alpha \Lambda^2
\label{scalarSE}
\end{equation}
In this case, there is no cancellation of quadratic singularities.
Thus if we expect our theory to be valid up to say $\Lambda \sim
10^{15\pm 5}$~GeV we would need an incredible fine tuning between
$m^2_S\simeq 10^4$~GeV$^2$ (which is the typical Higgs mass) and
$\delta m^2_S\simeq \alpha\, 10^{30\pm 10}$~GeV$^2$.  While this is
not inconsistent as such it is not what is expected, even more so as
this fine tuned cancellation would have to be repeated order by order
in perturbation theory. Thus we are nudged towards thinking that the
Standard Model may be valid only up to values of $\Lambda \sim
10^3$~GeV such that $m^2_S$ and $\delta m^2_S$ are of the same order.

What happens above $\Lambda \sim 10^3$~GeV?  This is where susy comes
into play. We have seen that in the fermion and gauge boson case it
was a symmetry that protected the masses from quadratic
singularities. In the case of the scalars this role can be played by
susy. For each fermion loop there are diagrams with a susy scalar
partner in the loop and adding them all up, the quadratic
singularities cancel. It is not surprising that this works. After all,
we have seen that fermionic masses are protected from quadratic
singularities in any theory. Since susy relates scalar masses to
fermion masses, in a susy theory scalar masses have to be protected as
well.

If there are susy partner particles, they should show up at about
$\sim 10^3$~GeV in order to be of any use in the solution of the
hierarchy problem. If there are susy partners at $\sim 10^3$~GeV an
unnaturally small tree-level value for the Higgs mass would be
protected from large radiative corrections.  However, there is still
no explanation, why the mass is small in the first place. This is
connected to another problem, the $\mu$-problem. Looking at
\Eqn{Wmssm}, the natural value for the couplings $y_u$, $y_d$ and
$y_e$ are ${\cal O}(1)$ and the natural value of $\mu$ would be of the
order of the largest scale, i.e. the cutoff scale where our theory
breaks down. We just convinced ourselves that thanks to the
cancellation of quadratic singularities this could well be $\sim
10^{15\pm 5}$~GeV. However, the term $\mu \, H_u\, H_d$ governs the
electroweak symmetry breaking, thus is of the order of the weak
scale. Finding an explanation (and there are many proposed) why $\mu$
is so small would be a solution to the $\mu$-problem.

When breaking susy by hand we want to make sure that we do not disturb
the cancellation of quadratic singularities. We call this {\it soft
  breaking} of susy. To get a rough understanding which terms are
allowed in soft breaking let us perform a simple dimensional
analysis. The correction to the scalar mass squared all have to have
mass dimension $[\delta m_S^2] = 2$. They can take the form
\begin{equation}
\delta m_S^2 = 
c^{(2)} \Lambda^2 + c^{(0)} \log(\Lambda/m_L)
\label{deltamsc}
\end{equation}
where $c^{(i)}$ is the product of the two couplings in the self-energy
diagram of the Higgs. Note that $[c^{(2)}]=0$ and $[c^{(0)}]=2$ and
that there is no term $c^{(1)} \Lambda$ as discussed in the fermion
case. We want to prevent terms of the form $c^{(2)} \Lambda^2$. This
can be done by allowing only susy breaking terms with mass dimension
strictly larger than 0, such that $[c]>0$. This eliminates the
possibility of a term $c^{(2)} \Lambda^2$ in \Eqn{deltamsc}. Thus we
can have scalar mass terms $-m_{ij}^2\, \varphi^\dagger_i\varphi_j$,
gaugino mass terms $-(m_{ij}/2)\ \lambda_i\lambda_j +{\rm h.c.}$ and
bilinear and trilinear scalar couplings $b_{ij}\, \varphi_i\varphi_j$
and $a_{ijk}\, \varphi_i\varphi_j\varphi_k$ as long as they respect
gauge invariance.

Writing down the general form of all terms with mass dimension equal
to or larger than one, taking into account the gauge symmetry of the
MSSM we get
\begin{eqnarray}
{\cal L}_{\rm soft} &=& 
 - \frac{1}{2} \Bigl(M_1\, \widetilde{B}\widetilde{B} + M_2\,
\widetilde{W}\widetilde{W} + M_3\, \tilde{g}\tilde{g} \Bigr) + 
{\rm h.c.} \nn \\
&-& m_{H_u}^2\,  h_u^\dagger\,  h_u - 
 m_{H_d}^2\, h_d^\dagger\,  h_d -
 ( b\,  h_u\,  h_d + {\rm h.c.}) \nn \\
&-& \Bigl({a_u} \tilde{u}_R \, \tilde{q}\,  h_u -
 {a_d}\, \tilde{d}_R\, \tilde{q}\, h_d -
  {a_e}\,\tilde{e}_R\, \tilde{l}\,  h_u \Bigr) + 
{\rm h.c.} \nn \\
&-& {m_Q^2}\, \tilde{q}^\dagger\,  \tilde{q}
 - {m_L^2}\, \tilde{l}^\dagger\, \tilde{l}
 - {m_u^2}\, \tilde{u}_R^\dagger\, \tilde{u}_R
 - {m_d^2}\, \tilde{d}_R^\dagger\, \tilde{d}_R
 - {m_e^2}\, \tilde{e}_R^\dagger\,  \tilde{e}_R
\label{MSSMsoft}
\end{eqnarray}
where ${a_i}$ and ${m_i^2}$ are $3\times 3$ matrices in family
space. Thus, the term ${m_u^2}\, \tilde{u}_R^\dagger\, \tilde{u}_R$
for example stands for $({m_u^2})_{f_i f_j}\,
(\tilde{u}_R^\dagger)^{f_i}\, (\tilde{u}_R)^{f_j}$. The doublets with
only the scalar component fields are denoted by $ h_u= (h_u^+,h_u^0)$
and $ h_d= (h_d^0,h_d^-)$ for the Higgs bosons and $\tilde
q=(\tilde{u}_L,\tilde{d}_L)$ and $\tilde l=(\tilde{\nu},\tilde{e}_L)$
for the matter multiplets. They are combined into $SU(2)$ gauge
invariant expressions as in \Eqn{Dmassterm}.

Note that there are no terms $m_{ij}\, \psi_i\psi_j+{\rm h.c.}$ and/or
$c_{ijk}\, \varphi_i^\dagger\, \varphi_j \varphi_k+{\rm h.c.}$ These
two terms are related since e.g. the first can be written as a
standard superpotential term plus $\varphi^\dagger\varphi$ and
$\varphi^\dagger \varphi\varphi +{\rm h.c.}$ terms. According to our
simple minded analysis they (or at least one of them) should be
present, since $[m_{ij}]= [c_{ijk}]= 1>0$.  However, a more careful
analysis~\cite{Girardello:1981wz} shows that these terms could lead to
quadratic divergences. It is remarkable that the soft terms can also
be explained as arising through a low energy effective theory of
spontaneously broken supergravity.

The terms in ${\cal L}_{\rm soft}$ give rise to additional
interactions, not listed in Eqs.~(\ref{phiFR})--(\ref{WphiFR}). The
scalar mass terms are not new as they have the same form as the first
interaction in \Eqn{WphiFR} and the others are given by
\begin{eqnarray}
- \frac{1}{2} \Bigl(M_1\, \widetilde{B}\widetilde{B} + M_2\,
\widetilde{W}\widetilde{W} + M_3\, \tilde{g}\tilde{g} \Bigr) + 
{\rm h.c.}
&\to&
\begin{picture}(100,30)(0,12)
\SetOffset(10,0)
\ArrowLine(0,15)(28,15)\Photon(0,15)(12,15){2}{1}\Photon(15,15)(26,15){2}{1}
\ArrowLine(60,15)(32,15)\Photon(34,15)(45,15){2}{1}\Photon(48,15)(60,15){2}{1}
\Text(30,15)[]{$\times$}
\end{picture}
\label{lambdaS}
\\
- ( b\,  h_u\,  h_d + {\rm h.c.})&\to&
\begin{picture}(100,30)(0,12)
\SetOffset(10,0)
\DashArrowLine(0,15)(22.5,15){2.5}
\DashArrowLine(45,15)(22.5,15){2.5}
\Text(22.5,15)[]{$\times$}
\end{picture}
\label{hhS}
\\
- \Bigl({a_u} \tilde{u}_R \, \tilde{q}\,  h_u -
 {a_d}\, \tilde{d}_R\, \tilde{q}\, h_d -
  {a_e}\,\tilde{e}_R\, \tilde{l}\,  h_u \Bigr) + 
{\rm h.c.}&\to&
\begin{picture}(100,30)(0,12)
\SetOffset(10,0)
\DashArrowLine(0,15)(25,15){3}
\DashArrowLine(45,30)(25,15){3}
\DashArrowLine(45,0)(25,15){3}
\Vertex(25,15){1.5}
\end{picture}
\label{uqhS}
\end{eqnarray}

\noindent
We have not depicted the hermitian conjugate of the various terms.

The cancellation of quadratic divergences discussed above is simply a
special case of so-called {\it non-renormalization theorems}. In fact,
in a perfectly supersymmetric theory there are even stronger
cancellations.  For example, it can be shown that the parameters of
the superpotential do not receive any quantum corrections at all.
Thus if a particle is massless at tree level, there will be no mass
generated at any order in perturbation theory, reminiscent of the
situation of gauge boson masses in gauge theories. Or if we choose a
(small) value of $\mu$ in the superpotential term $ \mu \, H_u\, H_d$,
susy protects this value to all orders. All there is is wave-function
renormalization of $\chi$SF and VSF, or alternatively gauge coupling
renormalization.  In these renormalization factors there are only
logarithmic singularities. There is a similar theorem related to the
Fayet-Iliopoulos term, \Eqn{FIterm}. If we set $k=0$ such a term will
not be generated at higher orders if we make the additional
requirement that the trace over the charges (associated with the
$U(1)$ under consideration) vanishes. As we have seen in
Section~\ref{sec:Dbreak} this can have important implications for the
breaking of susy.

\subsection{Towards the bigger picture} \label{sec:big}

The Lagrangian of the MSSM is given by \Eqn{masterL}, adapted to the
gauge group $SU(3)\times SU(2)\times U(1)$ with the superpotential
given in \Eqn{Wmssm} and supplemented by the soft breaking terms
\Eqn{MSSMsoft}. As the name suggests, in the unconstrained MSSM, there
are no constraints put on the soft breaking terms. This introduces a
large number of parameters. At the same time, for arbitrary values of
the parameters in \Eqn{MSSMsoft} we run very quickly in conflict with
experimental constraints. For example we get unacceptably large
flavour-changing neutral currents or large CP-violating effects (see
e.g. Ref~\cite{Martin:1997ns}).  If the terms in \Eqn{MSSMsoft} follow
a rather striking pattern in that they basically are proportional to
the Standard Model values (i.e. the matrices ${m_i^2}$ are
proportional to the identity matrix and ${a_i}$ are proportional to
the Yukawa coupling matrices) these dangerous terms are absent. This
also reduces the number of parameters drastically. A full $3\times 3$
matrix is replaced by a single parameter. We can go even further and
assume that some of these parameters are actually the same at some
high energy scale, leading to more and more constrained versions of
the MSSM.

The ultimate goal would be to understand the theory behind susy
breaking. In the top-down approach we start with a theory that is
valid up to very large scales. Such a theory then predicts all
soft-breaking parameters by considering the corresponding low-energy
effective theory. In this context, low-energy means TeV energy
scales. This would also have to include the gravitino, the susy
partner of the spin 2 graviton. We have not mentioned this at all in
this article because it leads to non-renormalizable theories and is
beyond the scope of this article, but the fact that local susy is
directly related to gravity is a strong hint that for a full
understanding of susy breaking, gravity might play an important role.

Alternatively, in the bottom-up approach we try to determine as many
parameters of the MSSM as precisely as possible, in the hope that this
will provide sufficient information to hint towards the theory that is
behind susy breaking. With the LHC about to start in earnest this
approach gains momentum. But first of all, we have to find at least
some of the susy partner particles. If you still hear the sentence
``susy is just around the corner'' in a few years from now, you most
probably wasted your time reading this article.

\subsection*{Acknowledgement}

It is a pleasure to thank Alan Martin for suggesting to write this
article and for his efforts and comments to make sure it was
(hopefully) kept on an accessible level.

\begin{appendix}
\section{Notation and Conventions} \label{App:Conventions}

Minkowski indices are denoted by $\mu, \nu, \kappa, \rho\ \ldots ;
\ (\mu \in \{0,1,2,3\}) \ $, spinor indices by $\alpha, \beta, \gamma,
\ad, \bd\ \ldots ;\ (\alpha \in \{1,2\}) \ $, two-component Grassmann
spinors by $\theta, \zeta, \xi, \thb, \zeb, \xib $ and Weyl spinors by
$\psi, \chi, \lambda$. The summation convention is always used.

metric: $g^{\mu\nu} = g_{\mu\nu} = {\rm diag}\{1,-1,-1,-1\} $

Pauli matrices: $\sigma^1=\mat{0}{1}{1}{0}; \
                 \sigma^2=\mat{0}{-i}{i}{0}; \
                 \sigma^3=\mat{1}{0}{0}{-1}$ 

Dirac spinor: $ \Psi = \left( \begin{array}{c} \psi_\alpha \\ \chib^\ad
\end{array} \right); \ \ $  
adjoint Dirac spinor:  $ \ol{\Psi} \equiv \Psi^+ \gamma_0 = 
\left(  \chi^\alpha \ \ \psib_\ad \right) $

Grassmann spinor: 
$ \theta^\alpha = \left( \begin{array}{c} \theta^1 \\ \theta^2
\end{array} \right); \ \  
\thb^\ad = \left( \begin{array}{c} \thb^1 \\ \thb^2
\end{array} \right) ;$

Antisymmetric $\epsilon$-tensor:
$\epsilon^{1\, 2} = - \epsilon^{2\, 1} =
- \epsilon_{1\,2} = \epsilon_{2\, 1} \equiv 1; \quad
\epsilon^{1\, 1} =\epsilon^{2\, 2} = 
\epsilon_{1\, 1} =\epsilon_{2\, 2}=0\, ;
$

\noindent
where $\theta^1$ etc. are Grassmann variables, i.e. anticommuting
c-numbers. The bar on $\psib_\ad$ as well as the dotted index denote
hermitian conjugation, i.e. $\psib_\ad = \left[\psi_\alpha\right]^\dagger$
and $\chi^\alpha = [\chib^\ad]^\dagger$.

The two component Weyl spinors $\psi_\alpha$ (left-handed) and
$\psib^\ad$ (right-handed) transform under Lorentz transformations as
follows:
\begin{eqnarray}
\psi_\alpha^\prime &=& M_\alpha^{\ \beta} \psi_\beta; \qquad \quad
\psib_\ad^\prime \ =\ (M^*)_\ad^{\ \bd} \psib_\bd \nn \\
\psi^{\prime \alpha} &=& \psi^\beta\, (M^{-1})_\beta^{\ \alpha} ;  \quad
\psib^{\prime \ad}\ =\ \psib^\bd\, (M^{* -1})_\bd^{\ \ad} 
\label{app:WeylM}
\end{eqnarray}
where $M = \exp( i \frac{\vec{\sigma}}{2} (\vec{\vartheta}-i
\vec{\varphi}))$ and $\vec{\vartheta}$ and $\vec{\varphi}$ are the
three rotation angles and boost parameters respectively. The indices
can be raised/lowered through the totally antisymmetric
$\epsilon$-tensor.  This holds for the two-component spinors as well
as for any Grassmann variables:
\begin{eqnarray}
&&
\psi_\alpha = \epsilon_{\alpha \beta} \psi^\beta; \ \
\psi^\alpha = \epsilon^{\alpha \beta} \psi_\beta; \ \
\psib_\ad = \epsilon_{\ad \bd} \psib^\bd; \ \
\psib^\ad = \epsilon^{\ad \bd} \psib_\bd;  \nn \\
&&
\theta_\alpha = \epsilon_{\alpha \beta} \theta^\beta; \ \
\theta^\alpha = \epsilon^{\alpha \beta} \theta_\beta; \ \
\thb_\ad = \epsilon_{\ad \bd} \thb^\bd; \ \
\thb^\ad = \epsilon^{\ad \bd} \thb_\bd; 
\label{app:raiselower}
\end{eqnarray}
we also have 
\begin{eqnarray}
&&
\epsilon^{\alpha \gamma} \epsilon_{\gamma  \beta} 
     = \delta^\alpha_{\ \beta}\, ; \qquad
\epsilon_{\alpha \gamma} \epsilon^{\gamma  \beta}  
     = \delta_\alpha^{\ \beta}\, ;
\nn \\
&& \epsilon^{\alpha\gamma}\epsilon_{\beta\delta} = 
\delta^\alpha_\delta \, \delta^\gamma_\beta - 
\delta^\alpha_\beta \, \delta^\gamma_\delta
\label{app:epsdelta}
\end{eqnarray}
The product of two Grassmann spinors is defined through
\begin{eqnarray}
&& \theta \zeta \equiv \theta^\alpha \zeta_\alpha = 
\theta^\alpha \epsilon_{\alpha \beta} \zeta^\beta = 
- \zeta^\beta \epsilon_{\alpha \beta} \theta^\alpha = 
\zeta^\beta \epsilon_{ \beta \alpha} \theta^\alpha = 
\zeta^\beta  \theta_\beta = \zeta \theta \nn \\
&& \thb \zeb \equiv \thb_\ad \zeb^\ad = - \zeb^\ad \thb_\ad = \zeb \thb
\end{eqnarray}
In particular, using $\theta_1 = - \theta^2, \theta_2 = \theta^1$ we have
\begin{eqnarray}
&& \theta \zeta = - \theta^1\zeta^2 + \theta^2\zeta^1 
= \theta_2\zeta_1 - \theta_1\zeta_2 \nn \\
&& \thb \zeb = + \thb_1\zeb_2 - \thb_2\zeb_1 
= - \thb^2\zeb^1 + \thb^1\zeb^2 
\end{eqnarray}
Using the Pauli matrices, we define 
\begin{equation} 
(\sigma^\mu)_{\alpha \ad} \equiv
\{ 1,\sigma^1,\sigma^2,\sigma^3 \}_{\alpha \ad}\, ;  \ \
(\sigmab^\mu)^{\ad\beta} \equiv
\{ 1,-\sigma^1,-\sigma^2,-\sigma^3 \}^{\ad\beta}\, ; \ \ 
\label{app:pauli}
\end{equation}  
Note that $\sigma^\mu$ has lower undotted-dotted indices, whereas
$\sigmab^\mu$ has upper dotted-undotted indices. These two set of
matrices are also related by
\begin{equation}
(\sigma^\mu)_{\alpha\bd } = 
\epsilon_{\bd\ad} \epsilon_{\alpha\beta} (\sigmab^\mu)^{\ad\beta }\, ;
\quad
(\sigmab^\mu)^{\ad\beta } = 
\epsilon^{\beta\alpha} \epsilon^{\ad\bd} (\sigma^\mu)_{\alpha\bd }\, ;
\label{app:sigmarel}
\end{equation}
The bar on $\sigma$ is a well established but maybe somewhat
misleading notation. We have $(\theta\sigma^\mu\zeb)^\dagger =
\zeta\sigma^\mu\thb \neq \zeta\sigmab^\mu\thb$ and
$\theta\sigma^\mu\zeb = - \zeta\sigmab^\mu\thb$.

The $\gamma$-matrices are defined as
\begin{equation}
\gamma^\mu \equiv \mat{0}{\sigma^\mu}{\bar{\sigma}^\mu}{0}; 
\quad \gamma^5 \equiv
i \gamma^0 \gamma^1 \gamma^2 \gamma^3 =
\mat{-1}{0}{0}{1} \, ;
\label{app:gamma}
\end{equation}
and have the usual commutation relations $\{\gamma^\mu, \gamma^\nu\} =
2 g^{\mu \nu}$ which is a simple consequence of the identity
\begin{equation}
(\sigma^\mu)_{\alpha \ad} (\sigmab^\nu)^{\ad\alpha} = 
{\rm Tr}(\sigma^\mu\sigmab^\nu) = 2 g^{\mu\nu} 
\label{app:tr}
\end{equation}
We also need
\begin{eqnarray}
&&
\left(\sigma^{\mu\nu}\right)_\alpha^{\ \beta} \equiv \frac{1}{4} 
   \left(\sigma^\mu\sigmab^\nu - \sigma^\nu\sigmab^\mu\right)_\alpha^{\ \beta}
\, ; \qquad
\left(\sigmab^{\mu\nu}\right)^{\ad}_{\ \bd} \equiv \frac{1}{4} 
   \left(\sigmab^\mu\sigma^\nu -
   \sigmab^\nu\sigma^\mu\right)^{\ad}_{\ \bd}
\, ;\label{App:sigmamunu}
\\
&&
\left(\sigma^\mu\sigmab^\nu + \sigma^\nu\sigmab^\mu\right)_\alpha^{\ \beta} = 
2\, g^{\mu\nu}\, \delta_\alpha^{\beta}
\, ; \qquad \quad
\left(\sigmab^\mu\sigma^\nu + \sigmab^\nu\sigma^\mu\right)^\ad_{\ \bd} = 
2\, g^{\mu\nu}\, \delta^\ad_{\bd}
\, ; 
\label{app:sigmaToG}
\\[5pt]
&&
{\rm Tr}\left(\sigma^\mu\sigmab^\nu\sigma^\rho\sigmab^\kappa\right) = 
2\left(g^{\mu\nu} g^{\rho\kappa} - g^{\mu\rho}g^{\nu\kappa} +
g^{\mu\kappa} g^{\nu\rho} + i\, \epsilon^{\mu\nu\rho\kappa}\right)
\label{app:TrFour}
\\[5pt]
&&
{\rm Tr}\left(\sigmab^\mu\sigma^\nu\sigmab^\rho\sigma^\kappa\right) = 
2\left(g^{\mu\nu} g^{\rho\kappa} - g^{\mu\rho}g^{\nu\kappa} +
g^{\mu\kappa} g^{\nu\rho} - i\, \epsilon^{\mu\nu\rho\kappa}\right)
\label{app:TrFourCC}
\end{eqnarray}
where we have $\epsilon^{0123}=+ 1$. Using 
$$ \Phi = \left( \begin{array}{c} \lambda_\alpha \\ \phib^\ad
\end{array} \right); \ \  
\ol{\Psi}  = \left(  \chi^\alpha \ \ \psib_\ad \right) 
$$ 
the Lorentz covariant expressions can be written in two-component
notation as follows:
\begin{eqnarray}
\ol{\Psi} \Phi &=& \chi \lambda + \psib \phib =
  \chi^\alpha \lambda_\alpha + \psib_\ad \phib^\ad \nn \\
\ol{\Psi} \gamma^5 \Phi &=&  - \chi \lambda + \psib \phib=
  -\chi^\alpha \lambda_\alpha + \psib_\ad \phib^\ad \nn \\
\ol{\Psi} \gamma^\mu \Phi &=& \chi \sigma^\mu \phib - \lambda
\sigma^\mu \psib =
  \chi^\alpha (\sigma^\mu)_{\alpha \ad} \phib^\ad - 
  \lambda^\alpha (\sigma^\mu)_{\alpha \ad} \psib^\ad  
\label{app:covariants}\\
\ol{\Psi} \gamma^\mu  \gamma^5 \Phi &=& 
 \chi \sigma^\mu \phib + \lambda\sigma^\mu \psib  =
 \chi^\alpha (\sigma^\mu)_{\alpha \ad} \phib^\ad +
  \lambda^\alpha (\sigma^\mu)_{\alpha \ad} \psib^\ad    \nn \\
\ol{\Psi} \gamma^\mu  \gamma^\nu \Phi &=& \chi \sigma^\mu \sigmab^\nu
\lambda + \psib \sigmab^\mu \sigma^\nu \phib =
 \chi^\alpha (\sigma^\mu)_{\alpha \ad} (\sigmab)^{\ad \beta} \lambda_\beta
+ \psib_\ad (\sigmab^\mu)^{\ad \beta} (\sigma^\nu)_{\beta \bd}\phib^\bd\nn 
\end{eqnarray}
With the help of 
\begin{equation}
\theta^\alpha \theta^\beta = 
- \frac{1}{2} \epsilon^{\alpha\beta} (\theta\theta)\, ; \qquad
\thb^\ad \thb^\bd = \frac{1}{2} \epsilon^{\ad\bd} (\thb\thb)\, ;
\label{app:epsthth}
\end{equation} 
we can derive the following frequently used identities:
\begin{eqnarray}
&& (\theta\sigma^\mu\thb)\, \theta^\alpha \sigma^\nu_{\alpha\ad}= 
\frac{1}{2}(\theta\theta)\, 
 \thb_\bd(\sigmab^\mu \sigma^\nu)^\bd_{\ \ad} \nn \\
&& (\theta\sigma^\mu\thb)\, (\sigma^\nu)_{\alpha\ad}\thb^\ad = 
\frac{1}{2} (\thb\thb) \,
(\sigma^\nu\sigmab^\mu)_\alpha^{\ \beta}\theta_\beta  \nn \\
&& (\theta \sigma^\mu \thb) (\theta \sigma^\nu \thb) = \frac{1}{2}
g^{\mu \nu}  (\theta\theta) (\thb\thb) 
\label{app:reshuffle}
\\
&& (\theta \zeta) (\theta \xi) = - \frac{1}{2} (\theta\theta)
(\zeta\xi)\nn \\ 
&& (\thb\zeb) (\thb \xib)  = - \frac{1}{2} (\thb\thb) (\zeb\xib) \nn
\end{eqnarray}
The last two identities are known as Fierz rearrangement formul\ae.

The derivatives with respect to a Grassmann variable are defined as
follows:
\begin{equation}
\partial_\alpha \equiv \frac{\partial}{\partial \theta^\alpha}\,; \quad
\partial^\alpha \equiv  \epsilon^{\alpha \beta} \partial_\beta\,; \quad
\pab_\ad \equiv \frac{\partial}{\partial \thb^\ad}\,; \quad
\pab^\ad \equiv  \epsilon^{\ad \bd} \pab_\bd\,; 
\label{app:gmder}
\end{equation}
Note that the rules for raising/lowering indices imply that
\begin{eqnarray}
&&
\partial_\alpha \theta^\beta= \delta_\alpha^\beta\,; \quad
\partial^\alpha  \theta_\beta= - \delta^\alpha_\beta\,; \quad
\pab_\ad \thb^\bd = \delta_\ad^\bd\,; \quad
\pab^\ad \thb_\bd = - \delta^\ad_\bd\,; \label{App:partial1} \\
&&
\partial^\alpha \theta^\beta = \epsilon^{\alpha \beta}\,; \quad
\partial_\alpha \theta_\beta = -\epsilon_{\alpha \beta}\,; \quad
\pab^\ad \thb^\bd = \epsilon^{\ad \bd}\,; \quad
\pab_\ad \thb_\bd = -\epsilon_{\ad \bd}\,; \quad
\label{App:partial2}
\end{eqnarray}
Furthermore, the derivatives also anticommute with other Grassmann
variables. For example
\begin{equation}
\partial_\alpha (\theta\theta) 
= (\partial_\alpha \theta^\beta)\theta_\beta 
   - \theta^\beta (\partial_\alpha \theta_\beta) 
= \delta_\alpha^{\ \beta} \theta_\beta 
   - \theta^\beta(- \epsilon_{\alpha\beta})
= \theta_\alpha +\epsilon_{ \alpha\beta} \theta^\beta  
= 2 \theta_\alpha
\label{App:partial3}
\end{equation}
and similarly
\begin{equation}
\partial^\alpha (\theta\theta) =  2 \theta^\alpha\,; \quad 
\pab_\ad (\thb\thb) = - 2 \thb_\ad\,; \quad 
\pab^\ad (\thb\thb) = - 2 \thb^\ad\,;
\label{App:partial4}
\end{equation}
The minus signs in Eqs.~(\ref{App:partial1}), (\ref{App:partial2}) and
(\ref{App:partial4}) seem strange at first, however they are required
if we insist on raising and lowering spinorial indices with the
$\epsilon$-tensor. As a consequence, the Taylor expansion in Grassmann
variables also have some unexpected minus signs. For infinitesimal
$\zeta$ and $\zeb$ we have
\begin{eqnarray}
\phi(\theta+\zeta) &=& \phi(\theta) + \zeta\partial\, \phi(\theta)
 + {\cal O} (\zeta\zeta) 
= \phi(\theta) + \partial\zeta\, \phi(\theta)
 + {\cal O} (\zeta\zeta) \\
\phi(\thb+\zeb) &=& \phi(\thb) - \zeb\pab\, \phi(\thb)
 + {\cal O} (\zeb\zeb)
= \phi(\thb) - \pab\zeb\, \phi(\thb)
 + {\cal O} (\zeb\zeb) \label{app:taylor}
\end{eqnarray}
where $\phi$ is an arbitrary function. Note that Taylor expansions in
Grassmann spinors terminate after the second term, since expressions
like $\zeta^\alpha\, \zeta\zeta = 0$.

Integration with respect to Grassmann variables is defined through
\begin{equation}
\int {\rm d}\theta^1 \theta^1  =
- \int \theta^1 {\rm d}\theta^1  =1\,; \qquad
\int {\rm d}\theta^1 = 0\,;
\label{app:integrate}
\end{equation}
thus, it is really the same as differentiation. We define
\begin{eqnarray}
{\rm d}^2 \theta &\equiv& -\frac{1}{4} \epsilon_{\alpha\beta}\, 
{\rm d}\theta^\alpha{\rm d}\theta^\beta \\
{\rm d}^2 \thb &\equiv& -\frac{1}{4} \epsilon_{\ad\bd}\,
{\rm d}\thb^\ad{\rm d}\thb^\bd
\end{eqnarray}
This has been arranged such that \Eqns{intf}{intd} hold.

Finally, our convention for the generators and covariant derivatives
are given in Eq.~(\ref{Qrep}), (\ref{QBrep}) and (\ref{Ddef}).  Thanks
to our conventions for the derivatives with respect to Grassmann
variables, \Eqn{app:gmder}, they satisfy $Q^\alpha = \epsilon^{\alpha
  \beta} Q_\beta$ etc. and, in particular, $\theta Q = Q \theta$.
They have the following anticommutation relations:
\begin{eqnarray}
&& \{Q_\alpha,D_\beta\} = \{\bar{Q}_\ad,\bar{D}_\bd\} =
\{Q_\alpha,\bar{D}_\bd\} = 0 \nn \\
&& \{Q_\alpha,Q_\beta\} = \{\bar{Q}_\ad,\bar{Q}_\bd\} = 0 \nn \\
&& \{D_\alpha,D_\beta\} = \{\bar{D}_\ad,\bar{D}_\bd\} = 0 
\label{app:algQD} \\
&& \{Q_\alpha,\bar{Q}_\ad\} = 2 i (\sigma^\mu)_{\alpha\ad}
\partial_\mu = 2 (\sigma^\mu)_{\alpha\ad} P_\mu \nn \\
&& \{D_\alpha,\bar{D}_\ad\} = - 2 i (\sigma^\mu)_{\alpha\ad}
\partial_\mu = - 2 (\sigma^\mu)_{\alpha\ad} P_\mu \nn
\end{eqnarray}

\section{Sample Calculations \label{app:sample}}

In this Appendix we present the details of some of the calculations
referred to in the main text.

\subsection{Anitcommutation relation \label{app:DD}}

As an example for an anticommutation relation we consider
$\{D_\alpha,\bar{D}_\ad\}$. Using the definitions \Eqn{Ddef} we would
expect to get four terms. However the expression reduces immediately
to two terms since $\{\partial_\alpha, \pab_\ad\} =
\{\sigma^\mu_{\alpha\bd} \thb^\bd, \theta^\beta
\sigma^\mu_{\beta\ad} \} = 0$, due to the Grassmann nature of
$\partial_\alpha$ and $\theta^\beta$ etc. Thus
\begin{eqnarray}
\{ D_\alpha,\bar{D}_\ad\} &=& 
\{\partial_\alpha, - i\,\theta^\beta \sigma^\mu_{\beta\ad}
\, \partial_\mu\} +
\{- i\,\sigma^\mu_{\alpha\bd} \thb^\bd
\, \partial_\mu,  \pab_\ad\} \nn \\
&=&- i\,\sigma^\mu_{\beta\ad}\, 
   \partial_\mu \{\partial_\alpha,\theta^\beta\} 
- i\,\sigma^\mu_{\alpha\bd}\, 
   \partial_\mu \{ \pab_\ad, \thb^\bd \}
\nn \\
&=&- i\,\sigma^\mu_{\beta\ad}\, \partial_\mu\, \delta_\alpha^\beta
- i\,\sigma^\mu_{\alpha\bd}\, \partial_\mu\, \delta_\ad^\bd
\nn \\
&=& - 2\, i\,\sigma^\mu_{\alpha\ad}\, \partial_\mu
\end{eqnarray}
where we used $\{\partial_\alpha,\theta^\beta\} = \partial_\alpha
\theta^\beta + \theta^\beta \partial_\alpha = [\partial_\alpha
  \theta^\beta] - \theta^\beta \partial_\alpha + \theta^\beta
\partial_\alpha = \delta_\alpha^\beta$. Here it is understood that in
the term $[\partial_\alpha \theta^\beta]$ the derivative acts only
within the brackets, but in all other terms it acts to everything on
the right. The minus sign in the product rule for differentiation is
as e.g. in \Eqn{App:partial3} due to the Grassmann nature of
$\theta^\beta$ and $\partial_\alpha$.

In a similar way we immediately get relations like
$\{\partial_\alpha,\thb^\bd\} = 0$ which have to be used to verify the
remaining relations listed in \Eqn{app:algQD}.

\subsection{Gauge transformation of $U_\alpha$ \label{app:Utrsf}}

Let us verify \Eqn{Utrsf}. Using the definition of $U_\alpha$,
\Eqn{Udefnonab}, and the gauge transformation of the VSF,
\Eqn{gaugetrsf}, we get
\begin{eqnarray}
U_\alpha  &\rightarrowtail& -\frac{1}{8\, g} \bar D \bar D\,
e^{-2 i\, g\, \Lambda} e^{-2 g\, V} e^{2 i\, g\, \Lambda^\dagger} 
D_\alpha\, e^{-2 i\, g\, \Lambda^\dagger}e^{2 g\, V}e^{2 i\, g\,
  \Lambda}
\nn \\
&=& 
-\frac{1}{8\, g} \bar D \bar D\,
e^{-2 i\, g\, \Lambda} e^{-2 g\, V} 
D_\alpha\,e^{2 g\,V}e^{2 i\,g\,\Lambda}
\nn \\
&=& 
-\frac{1}{8\, g}
e^{-2 i\, g\, \Lambda} \bar D \bar D\, e^{-2 g\, V}
\left(\left[D_\alpha\, e^{2 g\, V}\right] +  e^{2 g\, V}D_\alpha\right)
e^{2 i\,g\,\Lambda}
\nn \\
&=& e^{-2 i\, g\, \Lambda}\, U_\alpha\, e^{2 i\,g\,\Lambda} 
-\frac{1}{8\, g}
e^{-2 i\, g\, \Lambda} \bar D \bar D\,D_\alpha\, e^{2 i\,g\,\Lambda}
\label{app:Uone}
\end{eqnarray}
In the first step we used $D_\alpha\Lambda^\dagger = 0$ since
$\Lambda^\dagger$ is a RH$\chi$SF. In the second step we applied the
product rule for $D_\alpha$ and it is understood that in
$\left[D_\alpha\, e^{2 g\, V}\right]$ the derivative acts only within
the brackets. We also used $\bar D_\ad\Lambda = 0$. In the last step
we used the definition of $U_\alpha$. What remains to be done is to
show that the second term in the last line of \Eqn{app:Uone}
vanishes. To do this we use \Eqn{app:algQD} to anticommute $\bar D$
through to act on $e^{2 i\,g\,\Lambda}$. 
\begin{eqnarray}
\bar D \bar D\,D_\alpha\, e^{2 i\,g\,\Lambda} 
= - \bar D^\ad \bar D_\ad\,D_\alpha\, e^{2 i\,g\,\Lambda} &=&
- \bar D^\ad \{\bar D_\ad,D_\alpha\}\, e^{2 i\,g\,\Lambda} +
\bar D^\ad\,  D_\alpha \bar D_\ad\, e^{2 i\,g\,\Lambda} 
\nn\\
&=&
2 (\sigma^\mu)_{\alpha\ad}\bar D^\ad P_\mu \, e^{2 i\,g\,\Lambda} +0
\nn \\
&=&
2 (\sigma^\mu)_{\alpha\ad}P_\mu\bar D^\ad \, e^{2 i\,g\,\Lambda}
\nn \\
& =& 0
\label{app:Utwo}
\end{eqnarray}
In the second last step we used $[\bar D^\ad, P_\mu]=0$. This
completes the proof of \Eqn{Utrsf}. Note that in the abelian case
$U_\alpha$ commutes with $\Lambda$ (they do not contain non-commuting
generators of a non-abelian gauge theory) and thus we obtain the
result that $U_\alpha$ is gauge invariant.

Of course, the corresponding equations for $\bar U_\ad$ can be
obtained in a completely analogous way.

\subsection{$U_\alpha$ in terms of component fields \label{app:Ucomp}}

This calculation is somewhat tedious and can be tackled in several
ways. One option is to first make the computation in the abelian case
and verify \Eqn{Uabelian}. Then \Eqn{Unaexp} can be used to identify
the additional terms required in the non-abelian case.  Here, we
perform the calculation directly for the non-abelian case and verify
\Eqn{Unonabelian}. We will repeatedly use the fact that
$\theta_\alpha\, \theta\theta = \thb_\ad\, \thb\thb = 0$ and the
identities listed in \Eqns{app:epsthth}{app:reshuffle}.

We first write $e^{V} = 1+V+V^2/2$ with $V=V^a T^a$ and have to keep
in mind that we have to rescale all component fields at the end by a
factor $2\, g$ to obtain $e^{2g\, V}$. Using \Eqn{WZgauge} we find
\begin{eqnarray}
e^V &=& 1 +\theta\sigma^\mu\thb\, v_\mu(x)
+ i\, \theta\theta\, \thb\lambdab
 - i\, \thb\thb\, \theta\lambda
+  \frac{1}{2}\theta\theta\,\thb\thb\,  D
+\frac{1}{2}\theta\sigma^\mu\thb\, v_\mu\, 
\theta\sigma^\nu\thb\, v_\nu
\nn \\
&=& 1 +\theta\sigma^\mu\thb\, v_\mu(\bar y)
+ i\, \theta\theta\, \thb\lambdab
 - i\, \thb\thb\, \theta\lambda
+  \frac{1}{2}\theta\theta\,\thb\thb\,  
\left(D+\frac{1}{2} v^\mu v_\mu-i\, \partial^\mu v_\mu\right)
\label{app:exp1}
\end{eqnarray}
where we used $x^\mu = \bar y^\mu -i\, \theta\sigma^\mu\thb$ and in
the last line of \Eqn{app:exp1} all component fields are functions of
$\bar y^\mu$. The advantage of this is that $D_\alpha\, \bar y^\mu =
0$, thus the covariant derivative acts only as $\partial_\alpha$ on
the explicit $\theta$ appearing in \Eqn{app:exp1}. The change from
$x^\mu$ to $\bar y^\mu$ matters only in the one term where the
argument is explicitly indicated. The colour matrices are understood,
i.e. $\lambda = \lambda^a T^a$ etc. and $v^\mu v_\mu = T^a T^b\,
(v^a)^\mu v^b_\mu$. Acting with $D_\alpha$ we get
\begin{eqnarray}
D_\alpha\, e^V &=& 
(\sigma^\mu\thb)_\alpha\, v_\mu(\bar y)
+ 2\, i  \theta_\alpha\, \thb\lambdab(\bar y)
 - i \thb\thb\, \lambda_\alpha
+  \theta_\alpha\,\thb\thb\,  
\Big(D+\frac{1}{2} v^\mu v_\mu-i\,\partial^\mu v_\mu\Big)
\nn \\
&=& 
(\sigma^\mu\thb)_\alpha\, v_\mu(y)
+ i\, \thb\thb (\sigma^\mu\bar\sigma^\nu\theta)_\alpha\, 
  \partial_\nu v_\mu
+ 2\, i  \theta_\alpha\, \thb\lambdab(y)
- (\sigma^\nu \partial_\nu \lambdab)_\alpha\, \theta\theta\, \thb\thb
\nn \\
&&
\qquad - i\ \thb\thb\, \lambda_\alpha
+  \theta_\alpha\,\thb\thb\,  
\Big(D+\frac{1}{2} v^\mu v_\mu-i\,\partial^\mu v_\mu\Big)
\label{app:epx2}
\end{eqnarray}
where in the last step we have changed coordinates once more, this
time from $\bar y^\mu$ to $y^\mu = \bar y^\mu -2i\,
\theta\sigma^\mu\thb$ in order to be able to exploit $\bar D_\ad\,
y^\mu = 0$ in what follows. This affects only the terms where the
dependence on $\bar y^\mu$ or $y^\mu$ is explicitly given.  Before we
can act with $\bar D \bar D$ we have to multiply \Eqn{app:epx2} by
$e^{-V} = 1-V+\ldots$. The omitted terms have no effect. Also, we can
directly use $y^\mu$ since the change from $x^\mu$ to $y^\mu$ again
results only in vanishing terms. Thus, restoring the colour labels and
colour matrices, we obtain
\begin{eqnarray}
e^{-V^a T^a} D_\alpha\, e^{V^b T^b} &=& T^a\bigg(
(\sigma^\mu\thb)_\alpha\, v^a_\mu
+ i\, \thb\thb (\sigma^\mu\sigmab^\nu\theta)_\alpha\, 
  \partial_\nu v^a_\mu
+ 2\, i  \theta_\alpha\, \thb\lambdab^a
 \label{app:exp3} \\
&& \qquad
-\ (\sigma^\nu \partial_\nu \lambdab^a)_\alpha\, \theta\theta\,
\thb\thb
- i\ \thb\thb\, \lambda^a_\alpha
+  \theta_\alpha\,\thb\thb\,  (D^a - i\, \partial^\mu v^a_\mu) \bigg)
\nn \\
&& \hspace*{-3.5cm} +\ \frac{1}{2}\, T^a T^b \bigg(
\theta_\alpha\,\thb\thb\,   (v^a)^\mu v^b_\mu  
- \thb\thb\, (\sigma^\nu\sigmab^\mu\theta)_\alpha\, 
   v^a_\mu v^b_\nu
- i\,\theta\theta\,\thb\thb (\sigma^\mu\lambdab^b)_\alpha\, v^a_\mu
+ i\,\theta\theta\,\thb\thb (\sigma^\mu\lambdab^a)_\alpha\, v^b_\mu
\bigg)
\nn
\end{eqnarray}
The last two terms actually form a commutator $[T^a,T^b]=if^{abc} T^c$
if we reshuffle colour indices $a\leftrightarrow b$ in one of the
terms. This will give rise to the non-abelian covariant derivative
\Eqn{covder}. If we use \Eqn{app:sigmaToG} in the form
\begin{eqnarray}
\theta_\alpha\, (v^a)^\mu v^b_\mu  
-  (\sigma^\nu\sigmab^\mu\theta)_\alpha\, 
   v^a_\mu v^b_\nu &=&
\theta_\beta\, v^a_\mu v^b_\nu \left(
 \delta^\beta_\alpha\, g^{\mu\nu}
 -(\sigma^\nu\sigmab^\mu)_{\alpha}^{\ \beta}\right)
\nn \\
&=&\theta_\beta\, v^a_\mu v^b_\nu \, \frac{1}{2} \left(
\sigma^\mu\sigmab^\nu - \sigma^\nu\sigmab^\mu\right)_{\alpha}^{\ \beta}
\end{eqnarray}
we can also write the other two terms proportional to $T^aT^b$ as a
commutator and we start to recover the non-abelian field-strength
tensor \Eqn{Fnonabelian}. The same game has to be played to combine
the terms $i\, \thb\thb (\sigma^\mu\sigmab^\nu\theta)_\alpha\,
\partial_\nu v^a_\mu$ and $-i\, \theta_\alpha\,\thb\thb\, \partial^\mu
v^a_\mu$, proportional to $T^a$. After this, we simply act with $\bar
D \bar D$. Since all component fields are functions of $y^\mu$, this
implies to only act with $\bar\partial \bar\partial$ on the explicit
terms with $\thb$. Thus terms not containing $\thb\thb$ will vanish
and using $\bar\partial \bar\partial\, \thb\thb=-4$ we get
\begin{eqnarray}
\bar D \bar D\, e^{-V^a T^a} D_\alpha\, e^{V^b T^b} &=&
2i\, (\sigma^\mu \sigmab^\nu \theta)_\alpha\,
\left( T^a\,\partial_\mu v^a_\nu- T^a\,\partial_\nu v^a_\mu 
+ \frac{i}{2}\, [T^a,T^b] v^a_\mu v^b_\nu \right) 
\label{app:exp4} \\
&& \hspace*{-2cm}
+\ 4\, \theta\theta \left(T^a (\sigma^\mu \partial_\mu\lambdab^a)_\alpha
+ \frac{i}{2}\,[T^a,T^b](\sigma^\mu\lambdab^b)_\alpha\, v^a_\mu\right)
+ 4i\, T^a \lambda^a_\alpha - 4\, \theta_\alpha\, T^a D^a
\nn 
\end{eqnarray}
Restoring all factors (including the rescaling factor $2g$ for each
component field) we recover \Eqn{Unonabelian}. The corresponding
expression for $\bar U_\ad$ reads
\begin{equation}
\bar U^a_\ad = \frac{i}{2}\,
 \thb_\bd (\sigmab^\mu\sigma^\nu)^\bd_{\ \ad}\, F^a_{\mu\nu}
+\thb\thb\, (D_\mu \lambda^a)^\alpha \sigma^\mu_{\alpha\ad} 
-i\, \lambdab_\ad^a - \thb_\ad\, D^a
\label{app:UBnonabelian}
\end{equation}
To obtain the abelian expression we simply set $f^{abc} = 0$.

What remains to be done is to compute the Lorentz invariant product
$U^a U^a \equiv U_\alpha^a (U^a)^\alpha = \epsilon^{\alpha\beta}\,
U_\alpha^a U_\beta^a\, $ and take its supersymmetric $\theta\theta$
component. The only slightly tricky part is the $F^{\mu\nu}F_{\mu\nu}$
term. Using \Eqn{app:epsdelta}, results in two traces which are
evaluated with the help of Eqs.~(\ref{app:tr}), (\ref{app:TrFour}) and
(\ref{app:TrFourCC}). We find
\begin{eqnarray}
\lefteqn{\big[ U^a U^a\big]_{\theta\theta} } \nn \\ 
&=& -\ \frac{1}{8} \Big( {\rm Tr}(\sigma^\kappa\sigmab^\rho)
\, {\rm Tr}(\sigma^\mu\sigmab^\nu) - 
 {\rm Tr}(\sigma^\kappa\sigmab^\rho\sigma^\mu\sigmab^\nu)\Big)
F^a_{\kappa\rho}\, F^a_{\mu\nu} 
+ 2 i\, (\lambda^a \sigma^\mu D_\mu\lambdab^a) + D^a D^a
\nn \\
&=&
- \frac{1}{2}(F^a)^{\mu\nu}F^a_{\mu\nu}+
\frac{i}{4}\, \epsilon^{\mu\nu\kappa\rho}F^a_{\kappa\rho}\, F^a_{\mu\nu} 
+2i\,  (\lambda^a \sigma^\mu D_\mu\lambdab^a) + D^a D^a
\label{app:UU}
\end{eqnarray}
In a similar manner we get
\begin{equation}
\big[\bar U^a \bar U^a\big]_{\thb\thb} = 
- \frac{1}{2}(F^a)^{\mu\nu}F^a_{\mu\nu}-
\frac{i}{4}\, \epsilon^{\mu\nu\kappa\rho}F^a_{\kappa\rho}\, F^a_{\mu\nu} 
-2i\,  (D_\mu\lambda^a \sigma^\mu \lambdab^a) + D^a D^a
\label{app:UbUb}
\end{equation}
Thus the terms containing
$\epsilon^{\mu\nu\kappa\rho}F^a_{\kappa\rho}\, F^a_{\mu\nu}$ cancel in
the sum of the two terms \Eqns{app:UU}{app:UbUb} and adding all
prefactors we obtain the non-abelian version of \Eqn{FFabelian}.

\subsection{$\phi^\dagger\, \phi$ \label{app:term1}}

Let us verify \Eqn{WZdcomp} and compute $\phi^\dagger\,
\phi|_{\theta\theta\, \thb\thb}$, taking \Eqns{lhxsf}{rhxsf} as
input. Neglecting terms that do not contain two $\theta$ and two
$\thb$ spinors we get \begin{eqnarray}
\phi^\dagger\, \phi &=&
\big[\varphi^\dagger\big] \big[-\frac{1}{4}\, 
\theta\theta\, \thb\thb\, \partial^\mu\partial_\mu \varphi\big] +
\big[ \sqrt{2}\, \thb\psib\big] \big[\frac{i}{\sqrt{2}}\,
\theta\theta\, (\partial_\mu \psi\sigma^\mu\thb) \big] \nn \\
&+& \big[i\,\theta\sigma^\mu\thb\, (\partial_\mu \varphi)^\dagger\big]
\big[-i\,\theta\sigma^\nu\thb\, \partial_\nu \varphi \big]
+ \big[-\frac{i}{\sqrt{2}}\,
\thb\thb\, (\theta\sigma^\mu\partial_\mu \psib) \big]
\big[\sqrt{2}\, \theta\psi\big]  \nn \\
&+&\big[- \frac{1}{4}\, \theta\theta\,\thb\thb\, (\partial^\mu\partial_\mu
\varphi)^\dagger\big] \big[ \varphi\big] 
+\big[- \thb\thb\, F^\dagger\big] \big[- \theta\theta\, F \big]
\end{eqnarray}
We now use the reshuffling identities listed in \Eqn{app:reshuffle}. In
particular we make use of $\thb\psib\, (\partial_\mu
\psi\sigma^\mu\thb) = -\frac{1}{2}\, \thb\thb\, (\partial_\mu
\psi\sigma^\mu\psib)$ and $\theta\psi\, (\theta\sigma^\mu\partial_\mu
\psib) = -\frac{1}{2}\, \theta\theta\, (\psi\sigma^\mu\partial_\mu
\psib)$ and obtain
\begin{eqnarray}
\phi^\dagger\, \phi &=&\theta\theta\, \thb\thb \Big(
-\frac{1}{4} \varphi^\dagger 
\, \partial^\mu\partial_\mu \varphi -
\frac{i}{2}(\partial_\mu \psi\sigma^\mu\psib)
+ \frac{1}{2} (\partial^\mu \varphi)^\dagger
 \partial_\mu \varphi \nn \\
&& \qquad \quad +\ \frac{i}{2} (\psi\sigma^\mu\partial_\mu \psib) 
- \frac{1}{4}  (\partial^\mu\partial_\mu
\varphi)^\dagger \, \varphi 
+ F^\dagger  F  \Big)
\label{app:der1}
\end{eqnarray}
Finally, using integration by parts $(\partial^\mu\partial_\mu
\varphi)^\dagger \, \varphi = - (\partial^\mu \varphi)^\dagger
 \partial_\mu \varphi$ we obtain the result given in \Eqn{WZdcomp}.

\subsection{$\phi^\dagger\, e^{2 g\, V}\, \phi$ \label{app:term2}}

In this section we compute the gauge interaction terms between matter
fields and gauge fields in a non-abelian gauge theory, i.e. the third
term in \Eqn{susynonabelian}. We assume the $\chi$SF transform under a
certain representation of the gauge group and $T^a_{ij}$ are the
generators in this representation.  Expanding $e^{2 g\, V}$ in the
Wess-Zumino gauge we get
\begin{equation}
\left(e^{2g\, V^a T^a}\right)_{ij} = \delta_{ij} 
+2 g\, V^a  T^a_{ij} + 2 g^2\,  V^a V^b \, T^a_{ik} T^b_{kj}
\label{app:expexpand}
\end{equation}
and higher terms vanish due to the presence of terms $\theta^\alpha\,
\theta\theta=0$ or $\thb_\ad\, \thb\thb = 0$.

The insertion of the first term on the r.h.s. of \Eqn{app:expexpand}
has already been computed in Section~\ref{app:term1}. For the second
term on the r.h.s. of \Eqn{app:expexpand} we use \Eqn{WZgauge} as well
as \Eqns{lhxsf}{rhxsf}. Ignoring terms that do not contain two
$\theta$ and two $\thb$ spinors we get
\begin{eqnarray}
\lefteqn{\phi_i^\dagger\, 2g \, V^a T^a_{ij}\, \phi_j =} && \\
&& \big[\varphi_i^\dagger\big] 
\big[2g\, \theta\sigma^\mu\thb\, v_\mu^a T^a_{ij}\big]
\big[-i\, \theta\sigma^\nu\thb\, \partial_\nu\varphi_j\big]
+ \big[\sqrt{2}\, \thb\psib_i\big]
\big[2g\, \theta\sigma^\mu\thb\, v_\mu^a T^a_{ij}\big]
\big[\sqrt{2}\, \theta\psi_j\big] \nn \\
&+& \big[i\, \theta\sigma^\nu\thb\, (\partial_\nu\varphi_i)^\dagger
  \big]
\big[2g\, \theta\sigma^\mu\thb\, v_\mu^a T^a_{ij}\big]
\big[\varphi_j\big]
+ \big[\varphi_i^\dagger\big] 
\big[-i\, 2g\, \thb\thb\, \theta\lambda^a T^a_{ij}\big]
\big[\sqrt{2}\, \theta\psi_j\big] \nn \\
&+& \big[\sqrt{2}\, \thb\psib_i\big]
\big[i\, 2g\, \theta\theta\, \thb\lambdab^a T^a_{ij}\big]
\big[\varphi_j\big] 
+\big[\varphi^\dagger_i\big] 
\big[\frac{1}{2}\theta\theta\, \thb\thb\, D^a T^a_{ij} \big] 
\big[\varphi_j\big] \nn
\end{eqnarray}
Proceeding as in the derivation of \Eqn{app:der1} we get
\begin{eqnarray}
\lefteqn{\phi_i^\dagger\, 2g \, V^a T^a_{ij}\, \phi_j
\big|_{\theta\theta\, \thb\thb}
=} && \label{app:der2}
\\
&-& ig\, \varphi_i^\dagger\, v_\mu^a\, T^a_{ij}\, \partial^\mu\varphi_j
+ig\, (\partial^\mu\varphi_i)^\dagger\, v_\mu^a\, T^a_{ij}\, \varphi_j
+g\, \psi_j\sigma^\mu\psib_i \, v_\mu^a\, T^a_{ij}
\nn \\
&+&
ig\, \sqrt{2}\, \varphi_i^\dagger\, (\lambda^a\psi_j)\, T^a_{ij}
-ig\, \sqrt{2}\, \varphi_j\, (\lambdab^a\psib_i)\, T^a_{ij}
+ g \,\varphi_i^\dagger D^a\varphi_j \, T^a_{ij}
\nn
\end{eqnarray}
Finally we compute the insertion of the third term on the r.h.s. of
\Eqn{app:expexpand}. We get only one term proportional to
$\theta\theta\, \thb\thb$
\begin{eqnarray}
\phi_i^\dagger\, 2g^2 \, V^a T^a_{ik}\, V^b T^b_{kj}\, \phi_j
\big|_{\theta\theta\, \thb\thb}
&=& \big[\varphi_i^\dagger \big] \big[2 g^2\,  (\theta\sigma^\mu\thb)\,
  v_\mu^a\, T^a_{ik}\,  (\theta\sigma^\nu\thb)\, v_\nu^b\, T^b_{kj}\big]
\big[\varphi_j\big]\big|_{\theta\theta\, \thb\thb} \nn \\
&=& 
g^2\, \varphi_i^\dagger\, v_\mu^a (v^b)^\mu\, 
T^a_{ik}T^b_{kj} \, \varphi_j
\label{app:der3}
\end{eqnarray}
which represents a four-point interaction between the scalars and the
gauge bosons.

Combining Eqs.~(\ref{app:der1}), (\ref{app:der2}) and (\ref{app:der3})
we see that the various terms combine into gauge invariant parts. The
terms with $\partial_\mu \varphi$ and $\partial_\mu\psi_i$ combine to
$(D_\mu \varphi_i)^\dagger \, D^\mu\varphi_i$ and $\frac{i}{2}(\psi_i
\sigma^\mu (D_\mu \psib)_i - (D_\mu \psi)_i \sigma^\mu \psib_i)$
respectively, where the (gauge) covariant derivatives are given in
\Eqn{gaugeCovDer}. This leaves us with the $F^\dagger\, F$ term of
\Eqn{app:der1} (which is eliminated through its equation of motion as
discussed in Section~\ref{sec:WZ}) and the interaction terms $ig\,
\sqrt{2}\, T^a_{ij} (\varphi_i^\dagger\, \lambda^a\psi_j-
\varphi_j\, \lambdab^a\psib_i)$ of \Eqn{app:der2} which are
explicitly written in \Eqn{masterL}.

\end{appendix}

\end{document}